\documentclass[preprint,12pt]{./elsarticle}

\usepackage{lineno,hyperref}

\usepackage{graphicx}
\usepackage{subcaption}
\usepackage{amssymb}
\usepackage{amsfonts}
\usepackage{amsmath} 
\usepackage{amsbsy} 
\usepackage{amsthm} 
\usepackage{mathrsfs} 
\usepackage{epstopdf} 
\usepackage{dsfont} 

\newtheorem{Theorem}{Theorem}
\newtheorem{hypothesis}{Hypothesis}
\newtheorem{corollary}{Corollary}

\begin{document}

\begin{frontmatter}

\title{A Generalisation of the Capstan Equation and a Comparison Against Kikuchi and Oden's Model for Coulomb's Law of Static Friction\tnoteref{t1}}
\tnotetext[t1]{This work is based on a PhD thesis submitted to UCL in 2016 \citep{jayawardana2016mathematical}, and the initial research was funded by The Dunhill Medical Trust [grant number R204/0511] and UCL Impact Studentship.}

\author[mymainaddress]{Kavinda Jayawardana\corref{mycorrespondingauthor}}
\address[mymainaddress]{TEK Optima Research Ltd, Unit 10 Westcroft Buiness Park Oakdene Drive, Three Legged Cross, Wimborne BH21 6FQ}
\ead{kavjayawardana@tekoptimaresearch.com; zcahe58@ucl.ac.uk}
\cortext[mycorrespondingauthor]{Corresponding author}

\begin{abstract}
In this article, we extend the capstan equation to non-circular geometries. We derive a closed form solution for a membrane with a  zero Poisson's ratio (or a string with an arbitrary Poisson's ratio) supported by a rigid prism (at limiting-equilibrium case and in steady-equilibrium  case) or supported by a rigid general cone (at limiting-equilibrium case only). Our models indicate that the stress profile of the elastic body depends on the change in curvature of the rigid obstacle at the contact region. As a comparison, we extend Kikuchi and Oden's  model for Coulomb's law of static friction \cite{Kikuchi} to curvilinear coordinates, and conduct numerical experiments to examine the properties of Coulomb's law of static friction implied by Kikuchi and Oden's model in curvilinear coordinates and implied by our generalised capstan equation. Our numerical results indicate that for a fixed coefficient of friction, the frictional force is independent of the Young's modulus of the elastic body, and increasing the curvature, increasing the Poisson's ratio and decreasing the thickness of the elastic body increases the frictional force. Our analysis also shows that for given a tension ratio, the coefficient  of friction inferred by each model is different, indicating that the implied-coefficient of friction is model dependent.
\end{abstract}


\end{frontmatter}



\section{Introduction}

\begin{figure}[!h]
\centering
\includegraphics[width=0.5 \linewidth]{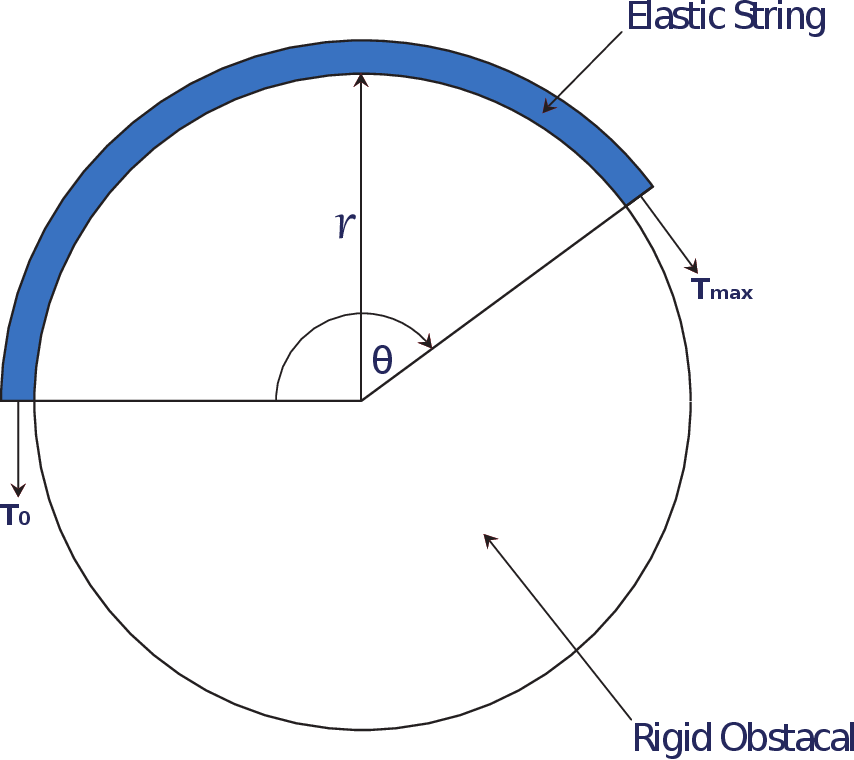}
\caption{A schematic representation of the capstan equation, where $T_0$ and $T_\text{max}$ are the minimum and the maximum tensions applied to the elastic string,  $r$ is the radius of the rigid obstacle and $\theta \in [0, \theta_\text{max}]$ is the angle of contact}
\end{figure}

In mathematical elasticity, the \emph{capstan equation} (also known as Euler's equation of tension transmission) is the relationship that governs the maximum applied-tension, $T_\text{max}$, and the minimum applied-tension, $T_0$, of an elastic \emph{string} (i.e. a one-dimensional elastic body, and by \emph{elastic} we do not mean the body is extensible in a mechanical sense), wound around a \emph{rough} (i.e. exhibiting friction, see chapter 13 of Johnson \cite{johnson1987contact} or section 5.2 of Quadling \cite{quadling2004mechanics}) rigid circular obstacle. Thus, the governing equation of this problem can be expressed as follows
\begin{align}
T_\text{max} = T_0 \exp(\mu_F\theta_\text{max}) , \label{CapstanEqn}
\end{align}
where $\theta_\text{max}$ is the contact angle and $\mu_F$ is the \emph{coefficient of friction} (see chapter 10 - problem 4 of Popov \cite{popov2010contact}), and where friction is defined as the force that opposes relative potential motion between the two bodies. The capstan equation is the most perfect example of a \emph{belt-friction model}, which describes behaviour of a belt-like object moving over a rigid obstacle subjected to friction \cite{rao2003engineering}. As a result of the model's simplicity, it is widely used to analyse the tension transmission behaviour of cable-like bodies in contact with circular profiled surfaces, such as in rope rescue systems, marine cable applications, computer storage devices (electro-optical tracking systems), clutch or brake systems in vehicles, belt-pulley machine systems and fibre-reinforced composites \cite{jung2008capstan}.\\

One of the largest applications of the capstan equation can be found in the field of electronic cable drive systems, and such devices include printers, photocopiers and tape recorders \cite{baser2010theoretical}. The most recent applications can be found in the field of robotics, as cable drive systems are fundamental in the design and manufacture of high-speed pick-and-place robots (DeltaBot, BetaBot and DashBot), wearable robot-assisted rehabilitation-devices (robotic prosthetics) \cite{kang2012design}, biologically-inspired humanoid-robots \cite{stellin2006preliminary}, and haptic devices \cite{ball2007planar,behzadipour2006cable,mustafa2008kinematic,perry2006design}. Further applications for the capstan equation can be found in the field of textiles, where tensioned fibers, yarns, or fabrics are frequently in contact with cylindrical bodies \cite{grosberg196919,jung2004effect}.\\ 

There exist cases in the literature where authors investigate the belt-transmission properties of noncircular pulley systems without the consideration of frictional interactions \cite{endo2010passive,kim2014design,kujawski2011analysis,shin2011variable,zheng2012non}. However, should one include friction for such cases then the ordinary capstan equation (\ref{CapstanEqn}) or most generalised capstan equations present in the literature (e.g. models by Baser and Konukseven \cite{baser2010theoretical}, Lu and Fan \cite{lu2013transmission}, Doonmez and Marmarali \cite{doonmez2004model}, Wei and Chen \cite{wei1998improved}, Jung \emph{et al.} \cite{jung2004effect,jung2008generalized,jung2008capstan}, Kim \emph{et al.} \cite{kim2004finite}, and Stuart \cite{stuart1961capstan}) will fail to be applicable, as these models assume that the rigid obstacle has a circular profile. Thus, to numerically model such problems, one may attempt to use Cottenden and Cottenden's \cite{cottenden2009analytical} model; however, Jayawardana \cite{jayawardana2016mathematical}  shows that Cottenden and Cottenden's work \cite{cottenden2009analytical} cannot be mathematically substantiated (see section 1.8 of Jayawardana \cite{jayawardana2016mathematical}). Another approach one may attempt is Konyukhov's \cite{konyukhov2015contact}, and Konyukhov's and Izi's \cite {konyukhov2015introduction} model for ropes (i.e. elastic strings) and orthotropic rough surfaces, where the authors use the \emph{generalised orthotropic friction law} by considering the coefficient of friction separately along the pulling and the dragging directions of a rope. Other generalisations of the capstan equation to non-circular geometries are derived by Maddocks and Keller \cite{maddocks1987ropes}, and Mack \emph{et al.} \cite{mack1953tension}: note that the latter publication more specifically deals with a rope wound around a cylinder at an angle.\\

As an alternative approach, we consider the belt-friction setting (i.e. elastic membranes over rigid obstacles) and generalise the capstan equation to be valid for rigid noncircular obstacles with zero Gaussian curvature, while still adhering to Coulomb's formulation of Amontons' laws of friction  (see section 10.2 of Popov \cite{popov2010contact}) in a \emph{static-equilibrium (or steady-equilibrium) dry-friction} setting (see section 11.3 of Kikuchi and Oden \cite{Kikuchi}).\\

Furthermore, given an observed tension pair $\{T_0, T_\text{max}\}$ from a real life experiment where the observations are made for the \emph{limiting-equilibrium} case, i.e. at the point of slipping (see section 5.1 Quadling \cite{quadling2004mechanics}), capstan equation (\ref{CapstanEqn}) can be used to calculate the coefficient of friction between the two bodies. However, should one use a different model (e.g. Kikuchi and Oden's model \cite{Kikuchi}, Konyukhov's model \cite{konyukhov2015contact}, etc.), one may find a different value for the coefficient of friction \cite{jayawardana2017quantifying}, despite the fact that the coefficient of friction is a materials property and should be independent of the friction model. For example, Kim \emph{et al.} \cite{kim2004finite} model a three-dimensional aluminium-sheet (assumed as an isotropic elastic-plastic material) deforming over a rough rigid cylinder with the finite-element method, where the coefficient of friction is calculated with the Wilson \emph{et al.}'s formulation \cite{wilson1991boundary}. From their numerical modelling, the authors find non-uniform pressure distributions at the contact region that is different to what is predicted by the standard capstan equation and an acquired (i.e. implied) coefficient of friction that is $7\%$ lower than the input coefficient of friction. Other comparative analysis include Zurek and Frydrych's work \cite{zurek1993comparative} on comparing capstan equation and Howell's equation (which is a power-law relationship between the frictional and normal forces) \cite{howell195324,howell195435} on measured yarn-on-yarn frictional forces. Thus, in this article, the model dependence of the coefficient of friction (capstan model vs Kikuchi and Oden's model) is also a subject of investigation.

\section{Generalising the Capstan Equation}

Consider an elastic body on a rough rigid surface that is subjected to external loadings and boundary conditions such that the body is at limiting-equilibrium. The governing equation for friction at the contact region can be expressed as follows
\begin{align}
F= \mu_FR , \label{FrictionLaw01}
\end{align}
where $R$ is the normal reaction force and $F$ is the frictional force experienced on the body, and $\mu_F$ is the coefficient of friction between the rough rigid surface and the body at the contact region (see equation 10.1 and 10.2 of Popov \cite{popov2010contact} or section 5.2 of Quadling \cite{quadling2004mechanics}). Equation (\ref{FrictionLaw01}) is Coulomb's mathematical interpretation of Amontons' laws of dry friction, and where the coefficient of friction experimentally found to be dependent on the pairing of the contacting materials, and in most cases independent of the contact area \cite{popova2015research}.\\

Note, hereafter, Einstein's summation notation (see section 1.2 of Kay \cite{kay1988schaum}) is assumed throughout, bold symbols signify that we are dealing with vector and tensor fields, and we regard the indices $i,j,k,l \in \{1,2,3\}$ and $\alpha,\beta,\gamma,\delta \in \{1,2\}$. Also, note that we usually reserve the vector brackets $\boldsymbol (\cdot \boldsymbol )_{\text{E}}$ for vectors in the Euclidean space and $\boldsymbol (\cdot \boldsymbol )$ for vectors in the curvilinear space. We refer the reader to Ciarlet \cite{ciarlet2005introduction} for a comprehensive study of elastic bodies in curvilinear coordinates as we adhere to most of the author's notations and conventions.\\

A \emph{true-membrane} is a planar elastic body (i.e. a 2-dimensional representation of a 3-dimensional thin elastic body with a constant thickness $h$) that is independent of any transverse effects (i.e. cannot admit any normal stresses or normal displacements), that does not resist bending and whose rest configuration is curvilinear (see chapter 7 of Libai and Simmonds  \cite{libai2005nonlinear}, and for \emph{shell-membranes} see section 4.5 of Ciarlet \cite{ciarlet2005introduction}). Now consider a true-membrane over a curvilinear rigid obstacle, where the contact region described by the surface $x^3=0$ in curvilinear coordinates $\boldsymbol(x^1, x^2, x^3\boldsymbol) \in \mathbb{R}^3$, where  $\mathbb{R}^k$ is the $k$-dimensional curvilinear space. Note, given that $\boldsymbol  T$ is the stress tensor of an elastic body, a true-membrane has the following properties: $\boldsymbol T = \boldsymbol  T (x^1, x^2)$, i.e. the stress tensor is a function of the planar coordinates only; $T^{3j} = 0$, $\forall j$, i.e. all normal components of the stress tensor are identically zero; and, for any constant values of $x^1$ and $x^2$, the only non-zero components of the stress tensor $T^{\alpha \beta}$ is constant $\forall x^3 \in [0,h]$, i.e. the planar stress components of the stress tensor are constant through the thickness of the membrane. Now, with some elementary tensor calculus, the friction law (\ref{FrictionLaw01}) can be expressed in the curvilinear space as $F_\alpha F^\alpha = (\mu_F)^2 F_3 F^3$.  Dividing this equation by the contact area twice is analogous to $\tau_3^\alpha\tau_\alpha^3 =  (\mu_F)^2 \tau^3_3\tau^3_3$, where $\tau_3^j = T_3^j|_{x^3=0}$ are normal stresses at the contact surface. As we are considering a true-membrane, the notion of normal stress becomes meaningless. However, note that the stress tensor is constant throughout the normal direction for such thin bodies, and thus, dividing our stress based friction equation further by the thickness of the body twice results in $f_{r\alpha} f^\alpha_r=  (\mu_F)^2 f_{r3} f^3_r$, where $f^\alpha_r$ is the frictional force densities and $f^3_r$ is the normal reaction force density of the true-membrane at the contact region. This leads to the following hypothesis:

\begin{hypothesis} \label{hyp1}
For a true-membrane over a rigid obstacle, whose contact area is described by the coordinates $\boldsymbol(x^1, x^2, 0\boldsymbol)$, at limiting-equilibrium, the force density field at the contact region is governed by the following equation
\begin{align}
\sqrt{f_{r\alpha} f^\alpha_r} = \mu_F \sqrt{f_{r3} f^3_r} \label{FrictionLaw02},
\end{align}
where $f_r^\alpha$ are the frictional force densities and $f_r^3$ is the normal reaction force density of the true-membrane in the curvilinear space, and $\mu_F$ is the coefficient of friction of the true-membrane and the rigid obstacle at the contact region.
\end{hypothesis}

\subsection{General Prism Case}

\begin{figure}[!h]
\centering
\includegraphics[width=1 \linewidth]{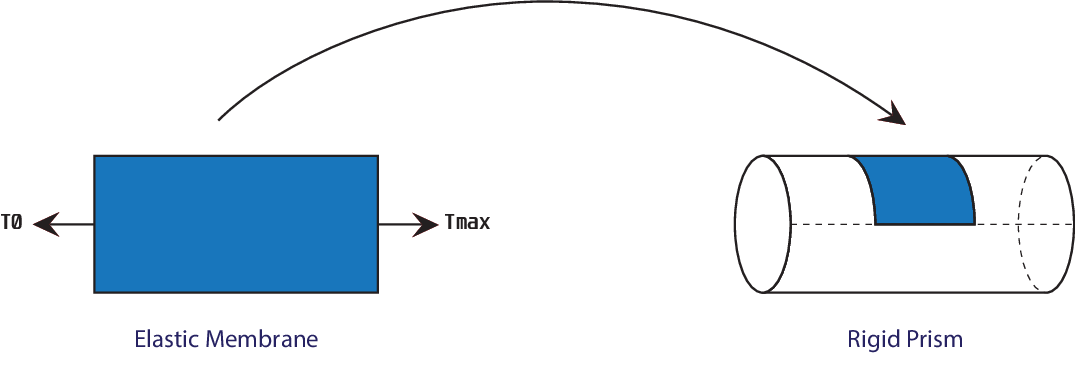}
\caption{A schematic representation of an elastic membrane over a rigid prism, where $T_0$ and $T_\text{max}$ are the minimum and the maximum tensions respectively applied to the elastic membrane in the azimuthal directions}
\label{genprism}
\end{figure}

Consider a general prism parametrised by the following  map
\begin{align*}
\boldsymbol\sigma(x^1,x^2) = \boldsymbol(x^1, ~u(x^2), ~v(x^2)\boldsymbol)_\text{E},~\forall~\boldsymbol(x^1, x^2\boldsymbol)\in\omega ,
\end{align*}
where $\omega \subset \mathbb{R}^2$ and $u(\cdot)$ and $v(\cdot)$ are $C^1(\omega)$ $2\pi$-periodic functions, $C^k(\cdot)$ is a space of continuous functions that have continuous first $k$ partial derivatives in the underlying domain. Note that $\omega$ is a simply-connected bounded two-dimensional domain with a positively-oriented piecewise-smooth closed boundary $\partial\omega$ such that $ \boldsymbol\sigma$ forms an injection (see section 1.3 of Pressley \cite{pressley2010elementary}). The prism's first fundamental form tensor can be expressed as $\boldsymbol{F}_{\!\text{[I]}} = \mathrm{diag}(1,~(u^\prime)^2+(v^\prime)^2)$ and the only non-zero component of its second fundamental form tensor can be expressed as follows
\begin{align*}
F_{\!\text{[II]}22} = - \frac{ \left(v^\prime u^{\prime\prime}-u^\prime v^{\prime\prime}\right)}{\sqrt{\left(u^\prime\right)^2 + \left(v^\prime\right)^2}} ,
\end{align*}
where the primes represent derivatives. Note that this is a surface with a zero Gaussian curvature, i.e. $\det(\boldsymbol F_{\!\text{[II]}}) = 0$ where $\det(\cdot)$ is the determinant of a matrix operator.\\

Now, consider a rectangular membrane with a zero Poisson's ratio (or a string with an arbitrary Poisson's ratio), with a thickness $h$ and a width $l$, that is in contact with the prism and at limiting-equilibrium (see Fig. \ref{genprism}) such that the boundary of this membrane can be described as follows
\begin{align*}
\partial \omega = \partial\omega_f\cup\partial\omega_{T_0}\cup\partial\omega_{T_\text{max}},
\end{align*}
where
\begin{align*}
\partial \omega_f & = \{\boldsymbol(x^1, ~x^2\boldsymbol) \mid x^1 \in\{0,l\}~ \text{and}~ \theta_0 < x^2 < \theta_\text{max}\} ~\text{(stress free boundary)},\\
\{\partial \omega_{T_0}, \partial \omega_{T_\text{max}}\}& = \{\boldsymbol(x^1, ~x^2\boldsymbol) \mid 0 < x^1 < l ~ \text{and} ~ x^2 = \{\theta_0,\theta_\text{max}\}\} 
\end{align*}
and where  $\theta_0 $ and  $\theta_\text{max}$ are the limits of the $x^2$ coordinate. The limits of $x^2$ is chosen such that $F_{\!\text{[II]}2}^{~~2}<0$ in $\omega$, i.e. the contact region is a surface with a positive mean-curvature, where the mean-curvature is defined as $H = -\frac{1}{2} F_{\!\text{[II]}\alpha}^{~~\alpha}$.\\

Consider the diffeomorphism $\boldsymbol \Theta (x^1,x^2,x^3) = \boldsymbol\sigma(x^1,x^2) + x^3 \boldsymbol{N}(x^1,x^2)$ with respect to the map $\boldsymbol\sigma$, where 
\begin{align*}
\boldsymbol N = \frac{\partial_1\boldsymbol \sigma \times \partial_2\boldsymbol \sigma}{||\partial_1\boldsymbol \sigma \times \partial_2\boldsymbol \sigma||}
\end{align*}
is the unit normal to the surface $\boldsymbol \sigma$, $\partial_\alpha$ are the partial derivatives with respect to the curvilinear coordinates $x^\alpha$, $\times$ is the Euclidean cross product and $||\cdot||$ is the Euclidean norm and $x^3 \in (-\varepsilon,\varepsilon)$ for some $\varepsilon>0$ (see theorem 4.1-1 of Ciarlet \cite{ciarlet2005introduction}). Now, with respect to the diffeomorphism $\boldsymbol \Theta$, the three-dimensional Cauchy's momentum equation in the  curvilinear space can be expressed as $ \nabla_{\!i} T^i_j + f_j = 0 $, where $\boldsymbol T$ is the Cauchy's stress tensor of the true-membrane,  $\boldsymbol f$ is a force density field (see section 1 subsection 3.4 of Morassi and Paroni \cite{Morassi}) and ${\boldsymbol \nabla}$ is the covariant derivative with respect to the curvilinear coordinate system $\boldsymbol ( x^1,x^2,x^3\boldsymbol)\in\mathbb{R}^3$ (see section 6.4 of Kay \cite{kay1988schaum}). By definition, $\boldsymbol f$ is the sum of all the force densities, and thus, one can re-express it as $\boldsymbol f = \boldsymbol f_r + \tilde{\boldsymbol g}_r$ where $\tilde{\boldsymbol g}_r$ is some external loading (e.g. effects due to gravity or centripetal force due to steady-equilibrium case), and $\tilde g_r^j$ are Lipschitz continuous (see definition (iii) of appendix A section 3 of Evans \cite{Evans}).\\

We consider boundary conditions of the following form
\begin{align}
T^1_\beta\big{|} _{\partial \omega} & = 0, ~\beta \in\{1,2\},  \label{PrismBC01x}\\
T^2_2\big{|} _{\partial \omega_{T0}} & = \frac{T_0}{hl}, \label{PrismBC01xx}\\
T^2_2\big{|} _{\partial \omega_{T\text{max}}} & = \frac{T_\text{max}\left(\mu_F,\boldsymbol \sigma(\omega), \tilde{\boldsymbol g}_r, T_0\right)}{hl}, \label{PrismBC01}
\end{align}
where $\theta_0<\theta_\text{max}$, and $T_0$ and $ T_\text{max}(\mu_F,\boldsymbol \sigma(\omega), \tilde{\boldsymbol g}_r, T_0)$ are forces applied at the boundary (minimum and maximum applied-tension respectively) such that $T_0< T_\text{max}(\mu_F,\boldsymbol \sigma(\omega), \tilde{\boldsymbol g}_r, T_0)$. Note that maximum applied tension is not arbitrary, as the maximum applied tension must have a very specific value for the membrane to remain at limiting-equilibrium, which depends on the minimum applied tension, the contact angle, the curvature, external loadings and the coefficient of friction. We further consider $g^1_r = 0$, and as the result of the zero Poisson's ratio and the boundary conditions, we find $f^1_r=0$, i.e. $f^1=0$.\\

Due to conditions, which include the zero Gaussian curvature, the zero Poisson's ratio and  $f^1=0$, and the construction of the boundary conditions (i.e. $x^1$ independence), we find that the only non-zero component of the stress tensor can be expressed as  $T^2_2 = T^2_2 (x^2)$. This result can be further justified as, even if one attempts this problem as a displacement-based problem in a linear elasticity framework, one comes to the same conclusion, i.e. if the stress tensor of the membrane is expressed as $T^\alpha_\beta = \frac{\nu E}{(1-\nu^2)} \epsilon^\gamma_\gamma(x^1, x^2) \delta^\alpha_\beta + \frac{ E}{(1+\nu)} \epsilon^\alpha_\beta(x^1, x^2)$, where $\boldsymbol{\epsilon}(\cdot)$, $E$ and $\nu$ are the is the strain tensor, Young's modulus and Poisson’s ratio of the membrane respectively (see section 4.5 of Ciarlet \cite{ciarlet2005introduction}) and  $\boldsymbol \delta $ is the Kronecker delta (see section 1.5 Kay  \cite{kay1988schaum}), then subjected to  zero Poisson's ratio, zero Gaussian curvature and boundary conditions (\ref{PrismBC01x}) - (\ref{PrismBC01}), one finds that the only non-zero component of the stress tensor can be expressed as $T^2_2 =  E \epsilon^2_2(x^2)$. Thus, the Cauchy's momentum equation at $x^3=0$ reduces to the following form
\begin{align*}
\partial_2 T^2_2 + f_{r2} + \tilde g_{r2} & = 0, \\
F_{\!\text{[II]}2}^{~~2} T^2_2 + f_{r3} + \tilde g_{r3} & = 0 .
\end{align*}

As friction opposes potential motion, i.e. $f_r^2<0$ ($f^2_r$ is decreasing as $x^2$ increases for our case), and the normal reaction force is positive, i.e. $f_r^3>0$ (as we are considering a unit outward normal to the surface), hypothesis \ref{hyp1} implies that $\sqrt{F_{\!\text{[I]}22}}_{~} f^2_r + \mu_F  f^3_r = 0$. This can be rearranged to express the following equation
\begin{align}
\partial_2 T^2_2 + \mu_F \sqrt{F_{\!\text{[I]}22}}_{~} F_{\!\text{[II]}2}^{~~2} T^2_2+ \tilde g_{r2} + \mu_F\sqrt{F_{\!\text{[I]}22}}_{~}\tilde g_{r3} = 0 .\label{PrismEqn02}
\end{align}

Finally, integrate equation (\ref{PrismEqn02}) (note that Lipschitz continuity of $\tilde g_r^j$ is necessary for the integration) with boundary condition (\ref{PrismBC01x}) and multiply the resulting solution by $lh$ to arrive at the following theorem:

\begin{Theorem}\label{thrmPrism}
The tension $T(\cdot)$ of a membrane with a zero Poisson's ratio (or a string with an arbitrary Poisson's ratio) in the azimuthal direction on a prism parametrised by the map $\boldsymbol(x^1, ~u(x^2), ~v(x^2)\boldsymbol)_\text{E}$, subjected to an external force field $\boldsymbol(0, ~g_r^2, ~g_r^3\boldsymbol)$ in the curvilinear space, at limiting-equilibrium is
\begin{align*}
T(\theta) = ~& \exp{ \left(- \mu_F\arctan\left(\frac{ v^\prime(\theta)} {u^\prime(\theta) }\right)\right)} \\
& \times \left[ C - \int^{\theta}_{\theta_0} \left(g_{r2}+\mu_F\sqrt{F_{\!\text{[I]}22}}_{~}g_{r3}\right) \exp{ \left(\mu_F\arctan\left(\frac{ v^\prime} {u^\prime}\right)\right)} d x^2\right],
\end{align*}
where $C= T_0 \exp(\mu_F\arctan(v^\prime /u^\prime)) |_{x^2=\theta_0}$, $T_0$ is the minimum applied-tension at $x^2=\theta_0$, $g_r^j$ are Lipschitz continuous, $u(\cdot)$ and $v(\cdot)$ are $C^1([\theta_0,\theta_\text{max}])$ $2\pi$-periodic functions and the interval $[\theta_0,\theta_\text{max}]$ is chosen such that $v^\prime u^{\prime\prime}-u^\prime v^{\prime\prime} >0$, $\forall ~x^2 \in [\theta_0,\theta_\text{max}]$ (i.e. the contact area has a positive mean-curvature with respect to its unit outward normal).
\end{Theorem}

\begin{proof}
Please see above for the derivation. Note that $\times$ is the scaler multiplication in this context. For the steady-equilibrium case, make the following transformation (i.e. if  $x \rightarrow y$, then replace $x$ with $y$, in this context)
\begingroup
\allowdisplaybreaks
\begin{align*}
g_{r2} & \rightarrow g_{r2} - \rho lh\Gamma^2_{\!22} V^2 ,\\
g_{r3} & \rightarrow g_{r3} +2 \rho lhH V^2,\\
\Gamma^2_{\!22} & =  \frac{ u^\prime u^{\prime\prime}+v^\prime v^{\prime\prime}}{\left(u^\prime\right)^2 + \left(v^\prime\right)^2},\\
H & = - \frac{1}{2}\frac{F_{\!\text{[II]}22}}{F_{\!\text{[I]}22}},
\end{align*}
\endgroup
where $\Gamma^2_{\!22}$ is a Christoffel symbols of the second kind (see section 6.3 of Kay \cite{kay1988schaum}), and $\rho$ is the mass density per unit volume, $l$ is the width, $h$ is the thickness and $V$ is the steady-speed of the membrane (in a Euclidean sense). This transformation follows from elementary tensor calculus (see section 11.3 of Kay \cite{kay1988schaum} and section 6.2 of Jayawardana \cite{jayawardana2016mathematical}).
\end{proof}

Note that theorem \ref{thrmPrism} is not valid when $u^\prime$ or $v^\prime$ are  zero between the limits $\theta_0$ and $\theta_\text{max}$. However, one can still evaluate such problems with some thought as $\arctan(x)$ remains finite in the limit $x \to \infty$ (such an example is examined in section \ref{examples}).\\

For a membrane with an arbitrary Poisson's ratio (where the range of the Poisson's ratio is $(-1,\frac{1}{2})$), we consider the following special case:

\begin{corollary}\label{crlPrism}
The tensile stress $\tau(\cdot)$ of a membrane, that is infinitely long in the $x^1$ direction and has an arbitrary Poisson's ratio in the azimuthal direction on a prism parametrised by the map $\boldsymbol(x^1, ~u(x^2), ~v(x^2)\boldsymbol)_\text{E}$ where $|x^1|\leq\infty$, subjected to an external stress field $\boldsymbol(0, ~\bar g_r^2, ~\bar g_r^3\boldsymbol)$ in the curvilinear space, at limiting-equilibrium is
\begin{align*}
\tau(\theta) = ~& \exp{ \left(- \mu_F\arctan\left(\frac{ v^\prime(\theta)} {u^\prime(\theta)}\right)\right)} \\
& \times\left[ C - \int^{\theta}_{\theta_0} \left(\bar g_{r2}+\mu_F\sqrt{F_{\!\text{[I]}22}}_{~}\bar g_{r3}\right) \exp{ \left(\mu_F\arctan\left(\frac{ v^\prime} {u^\prime}\right)\right)} d x^2\right],
\end{align*}
where $C= \tau_0 \exp{(\mu_F\arctan({ v^\prime} /{u^\prime}))} |_{x^2=\theta_0}$, $\tau_0$ is the minimum applied tensile stress at $x^2=\theta_0$,  $\bar g_r^j$ are Lipschitz continuous, $u(\cdot)$ and $v(\cdot)$ are $C^1([\theta_0,\theta_\text{max}])$ $2\pi$-periodic functions and the interval $[\theta_0,\theta_\text{max}]$ is chosen such that $v^\prime u^{\prime\prime}-u^\prime v^{\prime\prime} >0$, $\forall ~x^2 \in [\theta_0,\theta_\text{max}]$ (i.e. the contact angle has a positive mean-curvature with respect to its unit outward normal).
\end{corollary}

\begin{proof}
If the membrane in theorem \ref{thrmPrism} is infinitely long in $x^1$ direction, then any stresses along the $x^1$ direction will be independent of the stresses along the $x^2$ direction, implying that even if Poisson's ratio is not zero,  the stresses along directions $x^1$ and $x^2$ will be independent. Thus, for any cross-section where  $x^1=$ constant, we have our result.  For the steady-equilibrium case, make the following transformation
\begin{align*}
\bar g_{r2} & \rightarrow \bar g_{r2} - \rho \Gamma^2_{\!22} V^2 ,\\
\bar g_{r3} & \rightarrow \bar g_{r3} +2 \rho H V^2.
\end{align*}
\end{proof}

\subsection{General Cone Case}

\begin{figure}[!h]
\centering
\includegraphics[width=1 \linewidth]{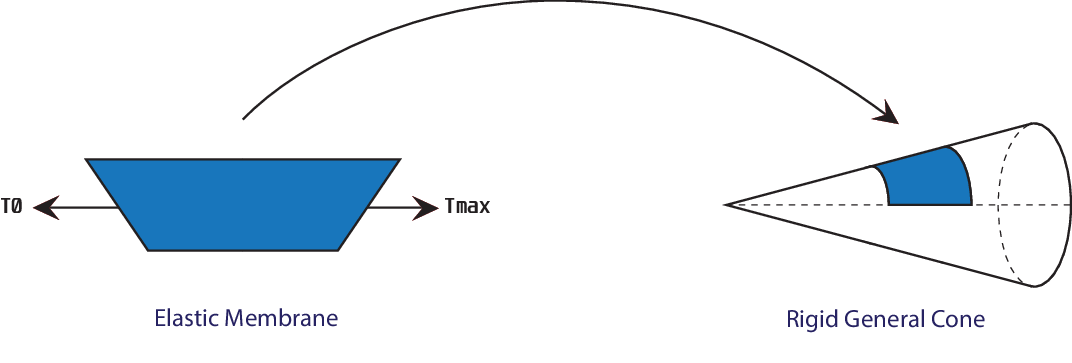}
\caption{A schematic representation of an elastic membrane over a rigid general cone, where $T_0$ and $T_\text{max}$ are the minimum and the maximum tensions respectively applied to the elastic membrane in the azimuthal directions. This image may be deceptive in its representation. The reader must understand that the parallel lines of the elastic membrane are following geodesics of the cone, and are \underline{\emph{not}} following the contours of the cone}
\label{gencone}
\end{figure}

Consider a general cone parametrised by the following map
\begin{align*}
\boldsymbol\sigma(x^1,x^2) = \boldsymbol(x^1, ~x^1\bar u(x^2), ~x^1\bar v(x^2)\boldsymbol)_\text{E},~\forall ~\boldsymbol(x^1, x^2\boldsymbol)\in\omega,
\end{align*}
where $\omega \subset \mathbb{R}_{>0}\!\times\!\mathbb R$, and $\bar u(\cdot)$ and $\bar v(\cdot)$ are $C^2(\omega)$ $2\pi$-periodic functions. Note that $\omega$ is a simply-connected bounded two-dimension domain with a positively-oriented piecewise-smooth closed boundary $\partial\omega$ such that $ \boldsymbol\sigma$ forms an injection. Thus, the cone's first fundamental form can be expressed as follows
\begin{align*}
\boldsymbol{F}_{\!\text{[I]}} = 
\begin{pmatrix}
1+\bar u^2+\bar v^2 & x^1\bar u\bar u^\prime+x^1\bar v\bar v^\prime \\
x^1\bar u\bar u^\prime+x^1\bar v\bar v^\prime & \left( x^1\bar u^\prime\right)^2+\left( x^1\bar v^\prime\right)^2
\end{pmatrix}.
\end{align*}
Also, the only nonzero component of the cone's second fundamental form can be expressed as follows
\begin{align*}
F_{\!\text{[II]}22} = -x^1 \frac{\left( \bar v^\prime \bar u^{\prime\prime}-\bar u^\prime \bar v^{\prime\prime}\right)}{\sqrt{\left(\bar u^\prime\right)^2 + \left(\bar v^\prime\right)^2 + \left(\bar v\bar  u^{\prime}- \bar u\bar v^{\prime}\right)^2}} .
\end{align*}

As a result of the non-diagonal nature of the first fundamental form of the cone, it is difficult to find a simple friction law as we did with the prism case. But note that this is a surface with a zero Gaussian curvature, i.e. $\det(\boldsymbol F_{\!\text{[II]}}) = 0$. Thus, Gauss' Theorema Egregium (see theorem 10.2.1 of Pressley \cite{pressley2010elementary}) implies that there exists a map $\boldsymbol\varphi: (\chi^1,\chi^2) \mapsto \omega$ such that the first fundamental form tensor with respect to the isometry $\boldsymbol\sigma\circ\boldsymbol\varphi$ is the $2\times 2$ identity matrix. With some calculations, we can define the properties of this map $\boldsymbol\varphi$ as follows
\begin{align*}
\chi^1 & = x^1\bar r \cos (\phi ),\\
\chi^2&= x^1 \bar r \sin (\phi ), 
\end{align*}
where
\begin{align*}
\bar r & = \sqrt{1+\bar u^2+\bar v^2}~,\\
\phi(x^2) & = \int^{x^2}_{0} \frac{\sqrt{\left(\bar u^\prime\right)^2 + \left(\bar v^\prime\right)^2+ \left(\bar v \bar u^{\prime}- \bar u\bar v^{\prime}\right)^2}}{\bar r^2} ~d\theta .
\end{align*}
Also,
\begin{align*}
\boldsymbol J =
\begin{pmatrix}
\bar r\cos(\phi)& x^1 \bar r^\prime\cos(\phi) - x^1 \bar r\phi^\prime \sin(\phi)\\
\bar r \sin (\phi)&  x^1\bar r^\prime \sin (\phi) +x^1 \bar r\phi^\prime\cos(\phi)
\end{pmatrix}.
\end{align*} 
is the Jacobian matrix of the map $\boldsymbol\varphi$. Furthermore, by the construction of  $\boldsymbol \varphi$ implies that $\det(\boldsymbol J) >0$, for all $ x^1 > 0$, i.e. $\boldsymbol \varphi$ is a valid unit-normal preserving coordinate transform for all $ x^1>  0$. With further calculations, one finds that the first fundamental form tensor of the cone with respect to the isometry $\boldsymbol\sigma\circ\boldsymbol\varphi$ is $F_{\!\text{[I]}\alpha\beta}^\varphi = \delta_{\alpha\beta}$, and that the second fundamental form tensor is can be expressed as follows
\begin{align*}
\boldsymbol{F}_{\!\text{[II]}}^\varphi = - \frac{\left(1+\bar u^2+\bar v^2\right)\left(\bar v^\prime \bar u^{\prime\prime}-\bar u^\prime \bar v^{\prime\prime}\right)}{x^1\left(\left(\bar u^\prime\right)^2 + \left(\bar v^\prime\right)^2 + \left(\bar v\bar  u^{\prime}- \bar u\bar v^{\prime}\right)^2\right)^{\frac32}}
\begin{pmatrix}
\sin^2(\phi) &-\sin(\phi) \cos(\phi)\\
-\sin(\phi) \cos(\phi) & \cos^2(\phi)
\end{pmatrix}.
\end{align*}

Now, consider an isosceles-trapezium membrane with a zero Poisson's ratio (or a string with an arbitrary Poisson's ratio), with a thickness $h$ and a width $l$ separating its parallel sides that is in contact with the cone such that it is at limiting-equilibrium (see Fig. \ref{gencone}) such that the boundary of this membrane has the following form
\begin{align*}
\partial \omega = \partial\omega_f\cup\partial\omega_{T_0}\cup\partial\omega_{T_\text{max}},
\end{align*}
where
\begin{align*}
\partial \omega_f &= \{(\chi^1,x^2) \mid\chi^1 \in \{d, d+l\} ~ \text{and}~ \theta_0 < x^2 < \theta_\text{max}\} ~\text{(stress free boundary)}, \nonumber\\
\{\partial \omega_{T_0}, \partial \omega_{T_\text{max}}\}&= \{(\chi^1,x^2) \mid d <\chi^1 < d+l ~ \text{and} ~ x^2 = \{\theta_0,\theta_\text{max}\}\}  ,
\end{align*}
and $d$ is the distance between the membrane and the apex of the cone at $x^2=0$. Note that $d$ must always be a positive constant, and, just as it is for the prism case, the limits of $x^2$ chosen such that $F_{\!\text{[II]}2}^{~~2}<0$ in $\omega$, i.e. the contact area is a surface with a positive mean-curvature.\\

Now, consider the diffeomorphism $$\boldsymbol\Theta(\chi^1,\chi^2, x^3) = \boldsymbol\sigma\circ\boldsymbol\varphi (\chi^1,\chi^2) + x^3 \boldsymbol{N}^\varphi(\chi^1,\chi^2)$$ with respect to the map $\boldsymbol\sigma\circ\boldsymbol\varphi $, where $\boldsymbol{N}^\varphi $ is the unit outward normal to the surface and $x^3 \in (-\varepsilon,\varepsilon)$, for some $\varepsilon>0$. Note that $ \boldsymbol{N}^\varphi(\chi^1,\chi^2) = \boldsymbol{N}(x^1,x^2)$, i.e. the unit normal to the surface is unchanged under the mapping $\boldsymbol\sigma\circ\boldsymbol\varphi $, and thus, the normal reaction force density remains unchanged under the new coordinate system. Now, with respect to the diffeomorphism $\boldsymbol\Theta$, the three-dimensional Cauchy's momentum equation in the curvilinear space can be expressed as $ \nabla_{\!i} T^i_j + f_j = 0 $ where is $\boldsymbol f$ is a force density field. Unfortunately, due to the geometry of the cone, we cannot impose a simple external-loading as we did with the prism case. Thus, we omit the external loading field $\boldsymbol g_r$ from the calculations, i.e. now we have $f^j = f_r^j$.\\

We consider boundary conditions of the following form
\begin{align}
T^1_\beta\big{|} _{\partial \omega} & = 0~, ~\beta \in\{1,2\}, \label{ConeBC010}\\
 T^2_2\big{|} _{\partial \omega_{T0}} & = \frac{T_0}{hl \cos(\phi(\theta_0))} , \label{ConeBC01}\\
T^2_2\big{|} _{\partial \omega_{T\text{max}}} & = \frac{T_\text{max}\left(\mu_F,\boldsymbol\sigma(\omega), T_0\right)}{hl\cos(\phi (\theta_\text{max}))}, \label{ConeBC012}
\end{align}
where $\theta_0<\theta_\text{max}$, and $T_0$ and $ T_\text{max}(\mu_F,\boldsymbol\sigma(\omega), T_0)$ are forces applied at the boundary such that $T_0< T_\text{max}(\mu_F,\boldsymbol\sigma(\omega), T_0)$. Comments that we made regarding $T_\text{max}(\mu_F,\boldsymbol\sigma(\omega), T_0)$ for the prism case applies to this case also, with the exception of the external loadings. As the result of the zero Poisson's ratio and the boundary conditions, we find $f^1_r=0$, i.e. $f^1=0$.\\

Due to conditions which include the zero Gaussian curvature, the zero Poisson's ratio and $f^1=0$, and the construction of the boundary conditions, we find that the only non-zero component of the stress tensor is $T^2_2 = T^2_2 (\chi^2)$. To illustrate this observation in more detail, recall that under the mapping $\boldsymbol\varphi$, the first fundamental form is the identity matrix (i.e. $F_{\!\text{[I]}\alpha\beta}^\varphi = \delta_{\alpha\beta}$) and the unit normal is preserved (i.e. $ \boldsymbol{N}^\varphi(\chi^1,\chi^2) = \boldsymbol{N}(x^1,x^2)$). This implies that the coordinates $(\chi^1, \chi^2)$ forms a \emph{geodesic} (or normal or rectangular) coordinate system (see chapter 10 of Berger \emph{et al.} \cite{berger2012differential}). Thus, for every constant value of $\chi^1$, the coordinate $\chi^2$ forms a \emph{geodesic}, i.e. a curve whose tangent vectors remain parallel if they are transported along it (see section 2.2 of Gavrilova \emph{et al.} \cite{gavrilova2016transactions}). This implies that  for every $c \in \mathbb{R}$, if $\chi^1 = c$, then $\chi^2 = c \tan(\phi) $ is a geodesic, and, similarly, for every $\tilde c \in \mathbb{R}$, if $\chi^2 = \tilde c$, then $\chi^1 = \tilde c \cot(\phi) $ is a geodesic. Note that a freely moving body always moves along a geodesic (see section 5.1 of Gasperini \cite{gasperini2013theory}). This implies that the stress along $\chi^2 = c \tan(\phi) $ will be independent of stress along $\chi^1 = \tilde c \cot(\phi) $, given that the membrane has a zero Poisson's ratio. As the boundary conditions are independent of $\chi^1$ (see equations (\ref{ConeBC010}) to (\ref{ConeBC012})), no friction is experienced along $\chi^1 = \tilde c \cot(\phi) $ geodesic, i.e. $f^1=0$, and thus, $T^2_2 = T^2_2 (\chi^2)$ is the only non-zero component of the stress tensor. Furthermore, the Cauchy's momentum equation at $x^3=0$ reduces to the following form
\begin{align*}
\partial_2 T^2_2 + f_{r2} & = 0 ,\\
F_{\!\text{[II]}2}^{\varphi 2} T^2_2 + f_{r3} & = 0.
\end{align*}

As friction opposes potential motion, i.e. $f_r^2<0$ ($f^2_r$ is decreasing as $\chi^2$ increases for our case), and the normal reaction force is positive, i.e. $f_r^3>0 $ (as we are considering a unit outward normal to the surface), hypothesis \ref{hyp1} implies that
\begin{align}
\partial_2 T^2_2 + \mu_F F_{\!\text{[II]}2}^{\varphi 2} T^2_2 = 0 .\label{ConeEqn01}
\end{align}

Despite the fact that equation (\ref{ConeEqn01}) provides one with a simple relation for friction, it is near impossible to integrate with respect to $\chi^2$. But notice that $\chi^2$ is related to $\chi^1$ by the following equation
\begin{align}
\chi^2 = \chi^1 \tan(\phi) . \label{ConeEqn02}
\end{align}
Thus, in accordance with Fubini's theorem (see theorem 3-10 of Spivak \cite{spivak2018calculus}), one may keep $\chi^1$ fixed and take the differential of equation (\ref{ConeEqn02}) to find the following
\begin{align}
d \chi^2 = \chi^1 \frac{\sqrt{\left(\bar u^\prime\right)^2 + \left(\bar v^\prime\right)^2+ \left(\bar v \bar u^{\prime}- \bar u\bar v^{\prime}\right)^2}}{1+ \bar u^2 +\bar v^2}\sec^2(\phi)~dx^2  .\label{ConeEqn03}
\end{align}
Now, with the use of equation (\ref{ConeEqn03}), one can express equation (\ref{ConeEqn01}) purely in terms of $x^2$ as follows
\begin{align}
\frac{\partial}{\partial x^2} \log(T^2_2) - \mu_F \frac{\sqrt{1 + \bar u^2 + \bar v^2} \left(\bar v^\prime \bar u^{\prime\prime} - \bar u^\prime \bar v^{\prime\prime}\right) }{\left(\bar u^\prime\right)^2 + \left(\bar v^\prime\right)^2 + \left(\bar v \bar u^{\prime} - \bar u\bar v^{\prime}\right)^2 }\cos(\phi) = 0. \label{ConeEqn04}
\end{align}
Finally, integrate equation (\ref{ConeEqn04}) with respect to boundary condition (\ref{ConeBC01}) and multiply the result by $hl \cos(\phi(\theta_0))$ to arrive at the following theorem:

\begin{Theorem}\label{thrmCone}
The tension $T(\cdot)$ of a membrane with a  zero Poisson's ratio (or a string with an arbitrary Poisson's ratio) in the azimuthal direction on a general cone parametrised by the map $x^1\boldsymbol \vartheta(x^2)$, where $\boldsymbol \vartheta(x^2)=\boldsymbol( 1, ~\bar u(x^2), ~\bar v(x^2)\boldsymbol)_\text{E}$, subjected to no external-loadings, at limiting-equilibrium is
\begin{align*}
T(\theta) = T_0\exp\left(\mu_F \int^{\theta}_{\theta_0} ||\boldsymbol \vartheta||\frac{ \boldsymbol \vartheta^{\prime\prime} \cdot (\boldsymbol \vartheta^\prime\times \boldsymbol \vartheta) }{||\boldsymbol \vartheta^\prime \times \boldsymbol \vartheta||^2}\cos(\phi) ~dx^2 \right),
\end{align*}
where 
\begin{align*}\
\phi(x^2) = \int^{x^2}_{0} \frac{||\boldsymbol \vartheta^\prime \times \boldsymbol \vartheta||}{||\boldsymbol \vartheta||^2} ~d\theta ,
\end{align*}
and where $T_0$ is the minimum applied-tension at $x^2=\theta_0$, $\bar u(\cdot)$ and $\bar v(\cdot)$ are $C^2([\theta_0,\theta_\text{max}])$ $2\pi$-periodic functions, and the interval $[\theta_0,\theta_\text{max}]$ is chosen such that $\boldsymbol \vartheta^{\prime\prime} \cdot (\boldsymbol \vartheta^\prime\times \boldsymbol \vartheta)\cos(\phi) >0$, $\forall ~x^2 \in [\theta_0,\theta_\text{max}]$ (i.e. the contact area has a positive mean-curvature with respect to its unit outward normal).
\end{Theorem}

\begin{proof}
Please see above for the derivation.
\end{proof}

\subsection{Explicit Solutions}
\label{examples}

Consider a rigid cylinder with radius $a$, parametrised by the map $\boldsymbol\sigma(x,\theta) = \boldsymbol(x, ~a\sin(\theta),~a\cos (\theta) \boldsymbol)_\text{E}$. Now, consider a membrane with a zero Poisson's ratio (or a string with an arbitrary Poisson's ratio) over the cylinder at limiting-equilibrium that is also subjected to the force of gravity $\boldsymbol{g} = \boldsymbol(0, ~0, ~-g\boldsymbol)_\text{E}$, where $ g$ is the acceleration due to gravity. Note that the covariant force density due to gravity in the curvilinear space with respect to the map $\boldsymbol\sigma$ can be expressed as $\boldsymbol{g}_r = \boldsymbol(0, ~a\varrho g\sin (\theta), ~- \varrho g \cos (\theta)\boldsymbol)$, where $\theta$ is the  acute angle that the vector $\boldsymbol (0,0,1\boldsymbol)_\text{E}$ makes with the vector $\boldsymbol (0,a,0\boldsymbol)$. Now, given that we are applying a minimum tension $T_0$ at $\theta_0 = 0$, theorem \ref{thrmPrism} implies that the tension observed on the membrane has the following form
\begin{align}
T_{g}(\theta) = ~&\left(T_0 - ahl\varrho g\frac{1-\mu_F^2}{1+\mu_F^2}\right)\exp(\mu_F \theta) \nonumber\\
& + ahl\varrho g \left(\frac{1-\mu_F^2}{1+\mu_F^2}\cos(\theta) + \frac{2\mu_F}{1+\mu_F^2}\sin(\theta)\right) .\label{capGrav}
\end{align}
The properties of equation (\ref{capGrav}) in the steady-equilibrium case is analysed in chapter 6 of Jayawardana \cite{jayawardana2016mathematical} and the most notable results can be found in section 2.3 of Jayawardana \emph{et al.} \cite{jayawardana2017quantifying}, along with a derivation and a numerical analysis of a general belt-friction model. Also, as the reader can see that if $g=0$, then equation (\ref{capGrav}) reduces to the ordinary capstan equation (\ref{CapstanEqn}).\\

\begin{figure}[!h]
\centering
\includegraphics[width=0.5 \linewidth]{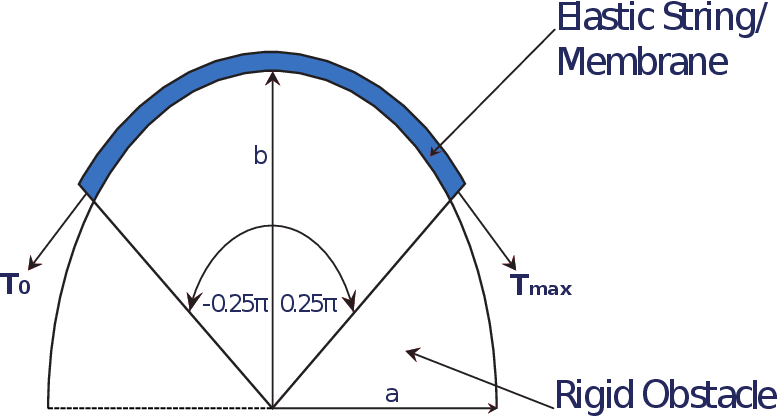}
\caption{A schematic representation of an elastic membrane over a rigid obstacle with an elliptical cross-section, where $a$ and $b$ are the respective horizontal and vertical radii of the rigid obstacle, and $T_0$ and $T_\text{max}$ are the minimum and the maximum tensions respectively applied to the elastic membrane in the azimuthal directions}
\label{ellipse}
\end{figure}

For an elliptical-prism case, consider a prism with a horizontal radius $a$ and a vertical radius $b$ (see Fig. \ref{ellipse}), parametrised by the map $\boldsymbol\sigma(x,\theta) = \boldsymbol(x, ~a\sin(\theta), ~b\cos (\theta) \boldsymbol)_\text{E}$, where $\theta$ is the  acute  angle that the vector $\boldsymbol (0,0,1\boldsymbol)_\text{E}$ makes with the vector $\boldsymbol (0,\varphi(\theta),0\boldsymbol)$, and where $\varphi(\theta) = (b^2\sin^2(\theta)+a^2\cos^2(\theta))^\frac12$. Now, consider a membrane with a  zero Poisson's ratio over the prism at limiting-equilibrium. Given that we are applying a minimum tension $T_0$ at $\theta_0 = 0$ and the membrane is not subject to an external loading, theorem \ref{thrmPrism} implies that the tension observed on the membrane has the following form
\begin{align}
T_{\text{elliptical}}(\theta) = T_0\exp\left(\mu_F \arctan\left(\frac{b}{a}\tan (\theta)\right)\right) \label{PrismEqn}.
\end{align}
Note that $\theta$ must not exceed the value $\frac{1}{2}\pi$, as at $\frac{1}{2}\pi$, as $\tan(\cdot)$ is singular. However, this is still not a problem as $\arctan((b/a)\tan (\cdot))$ remains finite at $\frac{1}{2}\pi$, i.e. $\arctan x \to \pm \frac{1}{2}\pi$ as $x \to \pm \infty$. For example, assume that the contact angle is $\frac{1}{2}\pi + \alpha$ where $0<\alpha<\frac{1}{2}\pi$. Thus, by considering the finiteness of $\arctan(\cdot)$ and considering some elementary trigonometric identities, one can find a solution of the following form
\begin{align*}
T_{\text{elliptical}}\!\left(\frac{1}{2}\pi+\alpha\right) = T_0\exp\left(\mu_F \left[\frac{1}{2}\pi + \arctan\left(\frac{b}{a}\tan (\alpha)\right)\right]\right) .
\end{align*}

\begin{figure}[!h]
\centering
\includegraphics[ width=0.75 \linewidth]{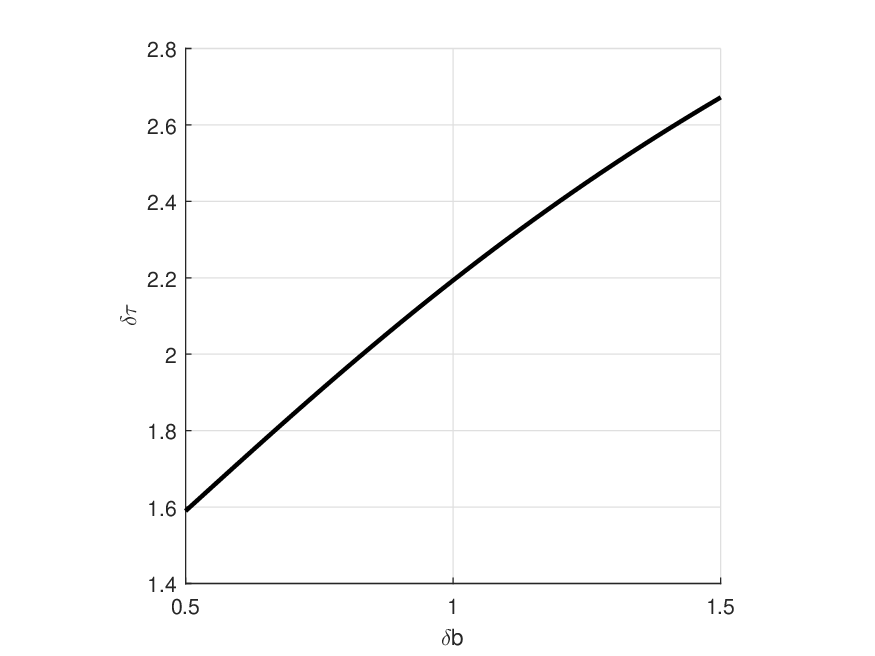}
\caption{Tension ratio (i.e. $\delta \tau$) against $\delta b$\label{PrismEqnPlot2}}
\end{figure}

Equation (\ref{PrismEqn}) implies that the maximum applied-tension, $T_\text{max}$, is dependent on the mean-curvature of the rigid prism. To investigate this matter further, consider a membrane (finite or infinitely long in $x^1$ direction) on a rough elliptical prism, at limiting-equilibrium, between the contact angles $\theta_0 = -\frac14\pi$ and $\theta_\text{max} = \frac14\pi$. Thus, theorem \ref{thrmPrism} (or corollary \ref{crlPrism}) implies the following
\begin{align}
\delta \tau = \exp\left(2\mu_F \arctan\left(\frac{b}{a}\right)\right) ,\label{PrismEqnPlot}
\end{align}
where $\delta \tau =\tau_\text{max}/\tau_0 = T_\text{max}/T_0$, which we call the \emph{tension ratio}. As the reader can see that for a fixed angle of contact and for a fixed coefficient of friction, equation (\ref{PrismEqnPlot}) implies a non-constant tension ratio, $\delta \tau$, for varying $\delta b$, where $\delta b= b/a$. As the mean-curvature of the prism is $ H(\theta) = \frac12 \frac{ab}{ (\varphi(\theta))^3}$, one can see that the tension ratio is related to the mean-curvature by the following equation
\begin{align*}
\delta \tau = \exp\left(2\mu_F \arctan\left[\max_{\theta \in [-\frac14\pi, \frac14\pi]}(2a H(\theta),1)+\min_{\theta \in [-\frac14\pi, \frac14\pi]}(2a H(\theta),1)-1\right]\right)  .
\end{align*}

We plot Fig. \ref{PrismEqnPlot2} to visualise the tension ratio against $\delta b$, which is calculated with $\mu_F = \frac12$ and $\delta b \in [\frac12, \frac32]$. The figure shows that for a fixed contact interval, as $\delta b$ increases (i.e. as the mean-curvature of the contact region increases), the tension ratio also increases. This is an intuitive observation as the curvature of the contact region increases, the normal reaction force on the membrane also increases, which in turn leads to a higher frictional force, and thus, a higher tension ratio. Now, this is a novel result as this effect cannot be observed with the ordinary capstan equation or other similar generalisations of the capstan equation in the literature \cite{maddocks1987ropes,cottenden2009analytical}.\\

Maddocks and Keller's work \cite{maddocks1987ropes} is considered to be the most comprehensive generalisation of the capstan equation to non-circular geometries in the literature. There, the authors postulate that the tension ratio of a rope that is wound around a rigid obstacle is independent of the curvature of the rigid obstacle as the path of the rope follows can always be re-parametrised by a different mapping, implying that the standard capstan equation holds true for non-circular geometries. This observation may be true for a complete turn of a rope around a rigid body (or $n$-number of complete turns, where $n$ is a natural number). However, theorem \ref{thrmPrism} (and corollary \ref{crlPrism}) implies that less than a complete turn (i.e. contact angle is less than $2\pi$ or not $2\pi n$), the tension ratio of the rope depends on the curvature of the rigid obstacle. Thus, the stress profile of the elastic body, i.e. $\delta \tau (\theta) = \exp\left(- \mu_F\arctan\left( v^\prime(\theta) /u^\prime(\theta)\right)\right)$, depends on the change in curvature of the rigid obstacle at the contact region. As far as we are aware, this is the first account of demonstrating this effect. It appears that this effect can also be predicted by Konyukhov's  work \cite{konyukhov2015contact}; however, no explicit statement regarding this is matter given by the author. \\

On a final note, to find the tension observed on a membrane over a right-circular cone with a $2\alpha$-aperture (i.e. $2\alpha$ is the angle between the two generatrix lines), consider the map $\boldsymbol\sigma(x,\theta) = \boldsymbol(x, ~x\tan(\alpha)\sin (\theta), ~x\tan(\alpha)\cos (\theta)\boldsymbol)_\text{E}$. Thus, in accordance with theorem \ref{thrmCone}, we find tension on the membrane can be expressed as follows
\begin{align*}
T_{\text{cone}}(\theta)  = T_0 \exp \left( \mu_F\cot[\alpha] \sin\left(\sin[\alpha]\theta\right) \right).
\end{align*}
In the limit $\alpha \to 0$ (i.e. membrane on an rigid cylinder with an infinitesimally small radius case) we find $T_{\text{cone}[\alpha = 0]}(\theta) = T_0 \exp (\mu_F \theta)$ (i.e. standard capstan equation), and in the limit $\alpha \to\frac{1}{2}\pi$ (i.e. membrane on a flat surface case) we find $T_{\text{cone}[\alpha = \frac{1}{2}\pi]}(\theta) = T_0$.

\section{A Comparison Against Kikuchi and Oden's Model for Coulomb's Law of Static Friction}

The most comprehensive mathematical study in friction that we are aware of is the publication by Kikuchi and Oden \cite{Kikuchi}, where the authors present a comprehensive analysis of the Signorini's problem, Coulomb's law of static friction and non-classical friction laws. The work includes meticulous documentation of the existence, the uniqueness and the regularity results for the given mathematical problems, which also includes finite-element modelling techniques. There, the authors present regularised Coulomb's law of static friction in Euclidean coordinates as follows
\begin{equation} \label{CoulombOden}
\qquad \sigma _T (\textbf{u}) = \left\{\begin{aligned}
\nu_F \sigma_n \frac{\textbf{u}_T}{|\textbf{u}_T|}  ,&~\text{if}~|\textbf{u}_T| \geq \varepsilon,\\
\nu_F \sigma_n \frac{\textbf{u}_T}{\varepsilon} ,~& ~\text{if}~|\textbf{u}_T| < \varepsilon,
\end{aligned}
\right.
\end{equation}
where $\nu_F$ is (again) the coefficient of friction, $\boldsymbol \sigma_T (\textbf u)$ is the normal-tangential stress tensor (i.e. shear), $\sigma_n < 0$ is the purely-normal stress (i.e. pressure), $\textbf{u}$ is the displacement field and $\textbf{u}_T$ is the tangential displacement field at the contact boundary, $\varepsilon>0$ is the regularisation parameter, and where this friction law (\ref{CoulombOden}) is considered to be an alternative mathematical representation of the Amontons' laws of dry friction. Recall that this model is defined for a Euclidean coordinate system. Thus, in this section, we modify Kikuchi and Oden's model for Coulomb's law of static friction \cite{Kikuchi}  to be valid in curvilinear coordinates, and we conduct several numerical examples and compare the results against our generalised capstan equation.\\

Let $\Omega\subset\mathbb{R}^3$ be a simply connected open bounded domain and let $\partial\Omega = \omega\cup\partial\Omega_f\cup\partial\Omega_T$ be the sufficiently smooth boundary of the domain, where $ \mathrm{meas}(\omega;\mathbb{R}^2)>0$, $\mathrm{meas}(\partial\Omega_f;\mathbb{R}^2)>0$ and $\mathrm{meas}(\partial\Omega_T;\mathbb{R}^2)>0$, and where $\mathrm{meas}(\cdot;\mathbb{R}^k)$ is standard Lebesgue measure in $\mathbb{R}^k$ (see chapter 6 of Schilling \cite{schilling2017measures}). Now, let ${\boldsymbol{X}}:\Omega\to \textbf{E}^3$ be a diffeomorphism where ${\boldsymbol{X}}(x^1,x^2,x^3) = \boldsymbol{\sigma} (x^1,x^2) + x^3\boldsymbol{N}(x^1,x^2)$, $\boldsymbol{\sigma} \in C^2(\omega;\textbf{E}^2)$ is an injective immersion and $\textbf{E}^k$ is the $k$-dimensional Euclidean space. Note that by construction, we either have $x^3>0$ or $x^3<0$  in $\Omega$, but never both.\\

Now, assume that $\Omega$ describes the domain of a curvilinear elastic body (i.e. whose rest configuration is curvilinear) such that $\omega$ describes the region where the elastic body is in contact with a rough rigid surface, $\partial\Omega_f$ describes the stress-free boundary and $\partial\Omega_T$ describes the boundary with traction. Let $\boldsymbol{v} \in C^2(\Omega;\mathbb{R}^3)$ be the displacement field of the body. Note that by the construction of domain $\Omega$, our elastic body resides either in the set $\boldsymbol{\sigma}(\omega) \times (0, h]$ (i.e. $x^3>0$ case) or the set $\boldsymbol{\sigma}(\omega) \times [-h, 0)$ (i.e. $x^3<0$ case) and choice depends on the problem one is considering. Given that $\boldsymbol{f}\in C^0(\Omega;\mathbb{R}^3)$ is an external force density field and $\boldsymbol{\tau}_0 \in C^0(\omega;\mathbb{R}^3)$ is a traction field (i.e. applied boundary-stress) at $\partial\Omega_T$, we can express the equations of equilibrium in curvilinear coordinates as  $\nabla_{\!i} T^i_j (\boldsymbol v) + f_j = 0$, where $T^{ij}(\boldsymbol v) = A^{ijkl}E_{kl}(\boldsymbol v) $ is the second Piola-Kirchhoff stress tensor, $ E_{kl}(\boldsymbol v) = \frac{1}{2}(\nabla_{\!i}v_j+\nabla_{\!j}v_i)$ is the linearised Green-St Venant strain tensor and $ A^{ijkl}= \lambda g^{ij} g^{kl} + \mu (g^{ik}g^{jl} + g^{il}g^{jk})$ is the isotropic elasticity tensor in curvilinear coordinates, $\lambda = \frac{\nu E}{(1+\nu)(1-2\nu)}$ is the first Lam\'{e}'s parameter, $\mu = \frac{1}{2}\frac{E}{(1+\nu)}$ is the second Lam\'{e}'s parameter, $E\in(0,\infty)$ is the Young's modulus and $\nu \in(-1,\frac12)$ is the Poisson's ratio of the elastic body, and where $g_{ij} = \partial_i {\boldsymbol{X}} \cdot \partial_j {\boldsymbol{X}}$ is the covariant metric tensor and $\cdot$ is the Euclidean dot product. The trivial boundary conditions are $ n_i T^i_j(\boldsymbol v)|_{\partial\Omega_f}= 0 $ and $ n_i T^i_j (\boldsymbol v)|_{\partial\Omega_T} = \tau_{0j}$, where ${\boldsymbol n}$ is the unit outward normal to $\partial\Omega$ and ${\boldsymbol \nabla}$ is the covariant derivative operator with respect to $\Omega$.\\

To investigate the behaviour at the boundary $\omega$, recall Kikuchi and Oden's  model for Coulomb's law of static friction \cite{Kikuchi}. Now, assume that ${\boldsymbol X}$ is constructed such that $v^3$ describes the normal displacement and $v^\beta$ describe the tangential displacements at the boundary $\omega$. Now, define the boundary as $\omega^\pm = \lim_{x^3\to 0^\pm}\Omega$ where the sign depends on the construction of $\Omega$. Noticing that on $\omega^-$, the form of the governing equations will remain identically to equation (\ref{CoulombOden}) and on $\omega^+$, we have $ T_3^3(\boldsymbol v)|_{\omega^+}=-\sigma_n (\boldsymbol v)$. Thus, we can re-express the friction equation (\ref{CoulombOden}) in curvilinear coordinates as follows
\begin{align}
v^3 |_{\omega^\pm} = 0,  \label{CoulombOden2}
\end{align}
\begin{equation} \label{CoulombOden3}
\qquad  T_3 ^\beta(\boldsymbol v) |_{\omega^\pm} = \left\{\begin{aligned}
\mp \left[\frac{\nu_F g_3  v^\beta}{(v_\alpha v^\alpha)^{\frac{1}{2}}}T_3^3(\boldsymbol v)\right] |_{\omega^\pm} & ,~\text{if}~(v_\alpha v^\alpha)^{\frac{1}{2}}|_{\omega^\pm} \geq \varepsilon ,\\
\mp \left[\frac{\nu_F g_3  v^\beta}{\varepsilon}T_3^3 (\boldsymbol v) \right]|_{\omega^\pm} &,~\text{if}~(v_\alpha v^\alpha)^{\frac{1}{2}} |_{\omega^\pm} < \varepsilon,
\end{aligned}
\right.
\end{equation}
where $\nu_F$ is the coefficient of friction, $T^3_3(\boldsymbol v)|_{\omega^\pm}$ is the purely-normal stress (i.e. normal reaction) and $T^\beta_3(\boldsymbol v)|_{\omega^\pm}$ are the normal-tangential stresses (i.e. shear) observed at the boundary $\omega^\pm$, $g_3 = \sqrt{g_{33}}|_{\omega} = 1$ by construction  and $(v_\alpha v^\alpha)^{\frac{1}{2}} = \sqrt{v_1v^1 + v_2v^2}$ by convention. \\

As $\boldsymbol T$ is the stress tensor of a curvilinear elastic body and it can be expressed as follows $$T^i_j(x^1,x^2,x^3) = \frac{\nu E}{(1+\nu)(1-2\nu)} E^k_k(x^1,x^2,x^3) \delta^i_j + \frac{E}{(1+\nu)} E^i_j(x^1,x^2,x^3),$$ and thus, modified Kikuchi and Oden's model explicitly depends on the Young's modulus, the Poisson's ratio and the curvature (the strain tensor explicitly depends on the curvature) of the elastic body, and implicitly depends on the thickness of the elastic body. Now recall our generalised capstan equation where the tension implied by theorem \ref{thrmPrism} can be expressed as $T(x^2) = h l E \epsilon^2_2(x^2)$ and the tensile stress implied by corollary \ref{crlPrism} can be expressed $\tau(x^2) = \frac{hE}{1-\nu^2}\epsilon^2_2(x^2)$ in a linear elasticity setting. This implies that capstan equation implicitly depend on Young's modulus, the Poisson's ratio and the thickness of the elastic body, and explicitly depends on the curvature (see equation (\ref{PrismEqn})) of the elastic body. This raises the question that what is the relationship between Kikuchi and Oden's  model Coulomb's law of static friction in curvilinear coordinates and the generalised capstan equation for different values of geometric and elastic properties?\\

\begin{figure}[!h]
\centering
\includegraphics[width=0.5 \linewidth]{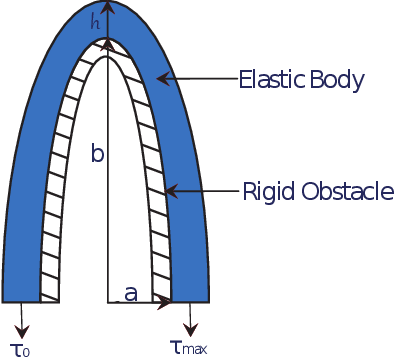}
\caption{A schematic representation of a curvilinear elastic body over a rigid obstacle with an elliptical cross-section, where where $a$ and $b$ are the horizontal and the vertical radii of the rigid obstacle, $h$ is the thickness of elastic body respectively, and $\tau_0$ and $\tau_\text{max}$ are the minimum and the maximum traction applied to the elastic membrane in the azimuthal directions}
\label{semiellipse}
\end{figure}

To investigate this matter, consider the following map of a rigid semi-prism,  $\boldsymbol (x^1, ~a\sin(x^2), ~b\cos(x^2)\boldsymbol)_\text{E}$, where $x^1 \in (-\infty,\infty)$, $x^2 \in [-\frac{1}{2}\pi,\frac{1}{2}\pi]$, $a$ is the horizontal radius and $b$ is the vertical radius (see Fig. \ref{semiellipse}). Now, assume that an elastic body is over this prism and one is applying a traction $\tau_0$ at $ x^2 =-\frac{1}{2}\pi$ and a traction $\tau_\text{max}$ at $ x^2 =\frac{1}{2}\pi$. Assume further that the elastic body has a constant  thickness $h$ and infinitely long in the $x^1$ direction. This leads to the following map of the unstrained configuration of the elastic body
\begin{align*}
{\boldsymbol{X}}(x^1,x^2,x^3) =  \boldsymbol(x^1, ~a\sin(x^2), ~b\cos(x^2)\boldsymbol)_\text{E} + \frac{x^3}{\varphi(x^2)}\boldsymbol(0, ~b\sin(x^2), ~a\cos(x^2)\boldsymbol)_\text{E},
\end{align*}
where $\varphi (x^2)= (b^2\sin^2(x^2)+a^2\cos^2(x^2))^{\frac{1}{2}}$, $x^1 \in (-\infty,\infty)$, $x^2 \in (-\frac{1}{2}\pi,\frac{1}{2}\pi)$ and $x^3 \in (0,h)$. Let $ \boldsymbol{v}= \boldsymbol{(}0,v^2(x^2,x^3),v^3(x^2,x^3)\boldsymbol{)}$ be the displacement field of the elastic body and let $ \delta\boldsymbol{v} = \boldsymbol{(}0,\delta v^2(x^2,x^3),\delta v^3(x^2,x^3)\boldsymbol{)}$ be a perturbation of the displacement field. Now, we can express the governing equations as follows
\begin{align*}
(\lambda + \mu)\partial_2( \nabla_{\!i} v^i ) + \mu (\bar\psi_2)^2  \Delta v^2 & = 0,\\
(\lambda + \mu)\partial_3( \nabla_{\!i} v^i ) + \mu  \Delta v^3 & = 0,\\
(\lambda + \mu)\partial_2( \nabla_{\!i} \delta v^i ) + \mu (\bar\psi_2)^2  \Delta \delta v^2 & = 0,\\
(\lambda + \mu)\partial_3( \nabla_{\!i} \delta v^i ) + \mu  \Delta \delta v^3 & = 0,
\end{align*}
where $\Delta = \nabla_i\nabla^i $ is the vector-Laplacian operator in the curvilinear space (see page 3 of Moon and Spencer \cite{moon2012field}) with respect to $\Omega^\text{New} = (-\frac{1}{2}\pi,\frac{1}{2}\pi) \times (0,h)$ and where $\bar\psi_2 = \varphi(x^2) + x^3 \frac{ab}{(\varphi(x^2))^2}$.\\

Eliminating the $x^1$ dependency, we can express the remaining boundaries as follows
\begin{align*}
\partial \Omega^{\text{New}} = \omega^{\text{New}} \cup \partial \Omega_f^{\text{New}} \cup \overline{\partial \Omega}_{T_0}^{\text{New}} \cup \overline{\partial \Omega}_{T_\text{max}}^{\text{New}},
\end{align*}
where
\begin{align*}
\omega^{\text{New}} & = (-\frac{1}{2}\pi,\frac{1}{2}\pi)\times \{0\} ,\\
\partial \Omega_f^{\text{New}} & = (-\frac{1}{2}\pi,\frac{1}{2}\pi)\times \{h\} ,\\
\partial \Omega_{T_0}^{\text{New}} &= \{-\frac{1}{2}\pi\}\times (0,h),\\
\partial \Omega_{T_\text{max}}^{\text{New}} &= \{\frac{1}{2}\pi\}\times (0,h),
\end{align*}
and where the overline represent closure of a set. Thus, the boundary conditions reduce to the following
\begingroup
\allowdisplaybreaks
\begin{align*}
v^3|_{\overline{\omega}^{\text{New}}} &= 0 ~\text{(zero-Dirichlet)} ,\\
\big[(\lambda + 2\mu)\left( \partial_2 v^2 + \Gamma^2_{\!22} v^2 + \Gamma^2_{\!23} v^3\right) + \lambda \partial_3 v^3\big]|_{\{\{\partial \omega_{T_0}^\text{New}, \partial \omega_{T_\text{max}}^\text{New}\}\times [0,h]\}} 
& = \{\tau_0 , \tau_\text{max}\}~\text{(traction)} ,\\
\big[(\bar\psi_2)^2\partial_3v^2 +\partial_2v^3\big]|_{\partial \Omega_f^{\text{New}}\cup\partial \Omega_{T_0}^{\text{New}} \cup \partial \Omega_{T_\text{max}}^{\text{New}}} & = 0~\text{(zero-Robin)} , \\
\big[\lambda\left( \partial_2 v^2 + \Gamma^2_{\!22} v^2 + \Gamma^2_{\!23} v^3\right) + (\lambda + 2\mu) \partial_3 v^3\big]|_{\overline{\partial \Omega}_f^{\text{New}}} & = 0 ~\text{(zero-Robin)} ,\\
\delta v^2|_{\overline{\partial\Omega}_f^{\text{New}}\cup\partial \Omega_{T_0}^{\text{New}} \cup \partial \Omega_{T_\text{max}}^{\text{New}}} &= 0 ,\\
\delta v^3|_{\partial \Omega^{\text{New}}} &= 0,
\end{align*}
\endgroup
where $ \Gamma^{2}_{\!22}$ and $ \Gamma^{2}_{\!23}$ are the  Christoffel symbols of the second kind.\\

Now we can find the friction laws governing the boundary conditions at the boundary $\overline{\omega}^{\text{New}}$, which can be expressed as follows:\\
If $\psi_2|v^2||_{\omega^{\text{New}}}\geq\epsilon$, then
\begin{align*}
\big[\mu \psi_2\partial_3v^2 + \nu_F \mathrm{sign}(v^2)T^3_3(\boldsymbol{v})\big]|_{\omega^{\text{New}}} = 0 ;
\end{align*}
If $\psi_2|v^2||_{\omega^{\text{New}}}<\epsilon$, then
\begin{align*}
\big[\mu \psi_2\partial_3\delta v^2 &+ \nu_F \epsilon^{-1} \psi_2v^2T^3_3(\delta \boldsymbol{v}) \\
&+ \nu_F \epsilon^{-1}\psi_2 \delta v^2 T^3_3 (\boldsymbol{v}) + \mu \psi_2\partial_3v^2 +\nu_F \epsilon^{-1}\psi_2v^2T^3_3(\boldsymbol{v}) \big]|_{\omega^{\text{New}}}= 0 ,
\end{align*}
where $ \psi_2 = \varphi(x^2)$ and $T^3_3(\boldsymbol v) =\lambda( \partial_2 v^2 + \Gamma^2_{\!22} v^2 + \Gamma^2_{\!23} v^3) + (\lambda + 2\mu) \partial_3 v^3$. Despite the fact that the original problem is three-dimensional, as a result of the problem's invariance in the $x^1$ direction, it is now a two-dimensional problem as the domain resides in the set $\{(x^2,x^3)\mid (x^2,x^3)\in [-\frac{1}{2}\pi,\frac{1}{2}\pi]\times[0,h]\}$.\\

To conduct numerical experiments, we use the second-order accurate iterative-Jacobi finite-difference method with Newton's method for nonlinear systems (see chapter 10 of Burden \emph{et al.} \cite{burden2015numerical}). Although we use a rectangular grid for discretisation, as a result of the curvilinear nature of the governing equations, there exists an implicit grid dependence implying that the condition $ \psi_0 \Delta x^2\leq \Delta x^3 $, $\forall ~ \psi_0 \in \{\bar\psi_2(x^2,x^3) \mid x^2\in[-\frac{1}{2}\pi,\frac{1}{2}\pi] ~\text{and}~ x^3\in[0,h] \}$ must be satisfied, where $\Delta x^j$ is a small increment in the $x^j$ direction in this context. For our purposes, we use $\Delta x^2 = \frac{1}{N-1}\pi$ and $\psi_0 = \bar\psi_2(\frac{1}{4}\pi,h)$, where $N = 250$. Also, for all our examples, we fix the values $\nu_F = \frac{1}{2}$, $\tau_0 =1$ (units: $\text{N}/\text{m}^2$), $a =2$ (units: m) and $\varepsilon = 10^{-5}$ (units: m) as our constants, and we assume the default values $\tau_\text{max}= 2$ (units: $\text{N}/\text{m}^2$), $b=2$ (units: m), $h=1$ (units: m), $ E = 10^3$ (units: Pa) and $\nu = \frac{1}{4}$ and  for our variables, unless it strictly says otherwise. Furthermore, all numerical codes are available at \href{http://discovery.ucl.ac.uk/id/eprint/1532145}{http://discovery.ucl.ac.uk/id/eprint/1532145}.

\begin{figure}[!h]
\centering
\includegraphics[trim = 2cm 1cm 2cm 1cm, clip = true, width=1\linewidth]{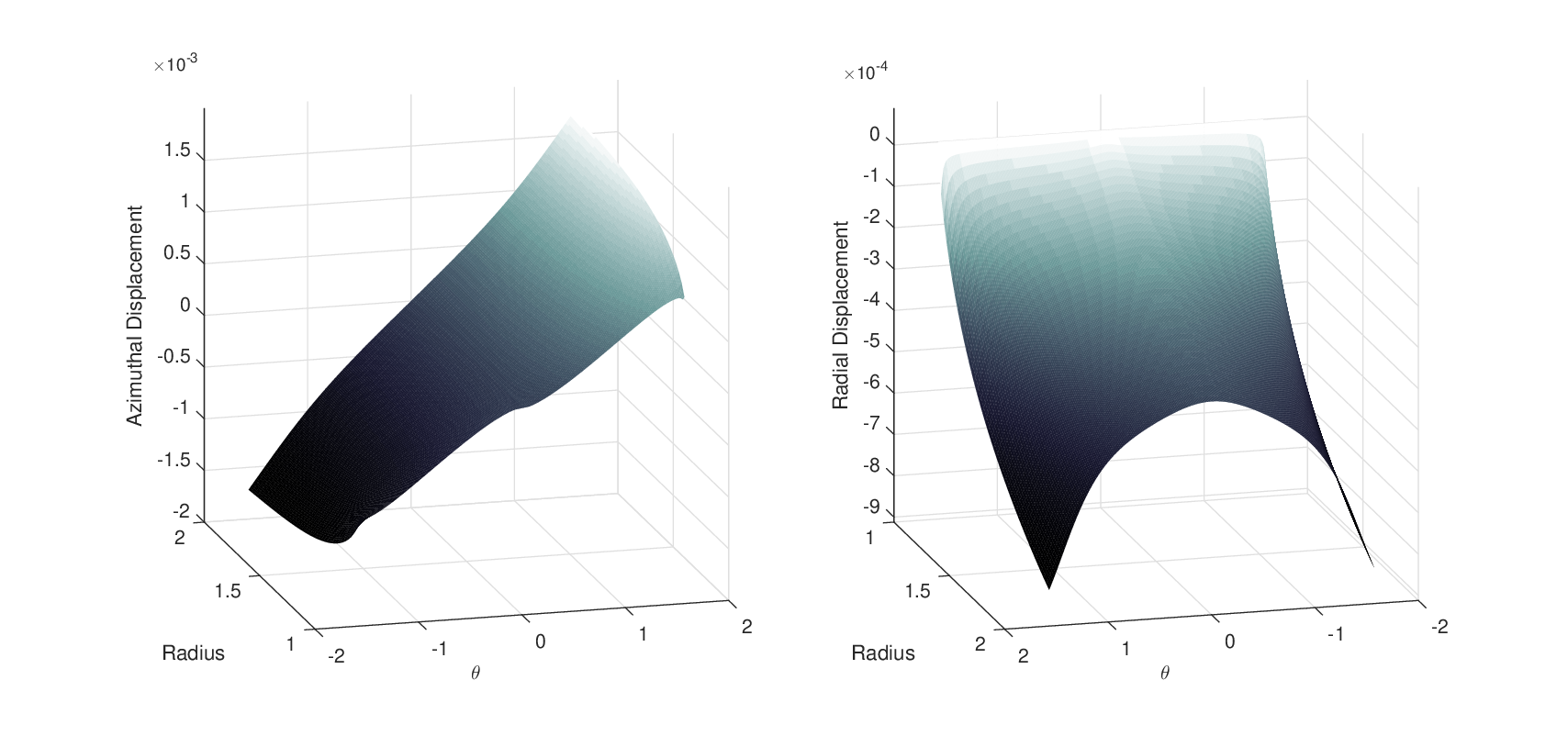}
\caption{Displacement field of the modified Kikuchi and Oden's model, where $\theta =x^2$ \label{Oden1Eg1}}
\end{figure}

\begin{figure}[!h]
\centering
\includegraphics[ width=0.625\linewidth]{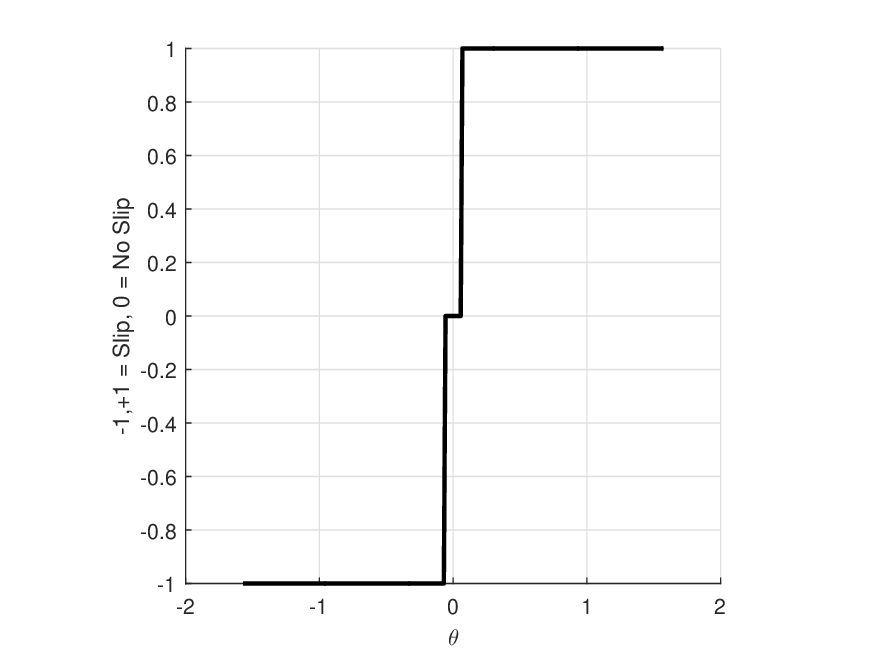}
\caption{Slip and stick regions of the modified Kikuchi and Oden's model, where $\theta =x^2$ \label{Oden1Eg2}}
\end{figure}

Fig.s \ref{Oden1Eg1} and \ref{Oden1Eg2} are calculated with the values $\tau_\text{max}= 1$ with a grid of $250\times 41$ points (i.e. a discretised domain with $250$ grid points in $x^1$ direction and $41$ grid points in $x^3$ direction). Fig. \ref{Oden1Eg1} shows the azimuthal (i.e $v^2$) and the radial (i.e $v^3$) displacements. The maximum azimuthal displacements are observed at $x^2=\pm\frac{1}{2}\pi$, with respective azimuthal displacements of $v^2 = \pm1.72\times 10^{-3}$rad. The maximum radial displacement is observed at $x^2=\pm\frac{1}{2}\pi$, with a radial displacement of $v^3 =-8.24\times 10^{-4}$m. Also, Fig. \ref{Oden1Eg2} shows the behaviour of the elastic body at the boundary $\overline{\omega}^\text{New}$. It implies that in the region $[-\frac{1}{2}\pi,-0.0694]$ the body slid in the negative (i.e. decreasing) azimuthal direction, and in the region $[0.0694,\frac{1}{2}\pi]$ the body slid in the positive (i.e. increasing) azimuthal direction. The region $(-0.0201, 0.0201)$ describes the azimuthal region of the body that is bonded to the rigid obstacle, i.e. magnitude of the displacement is  below the threshold of sliding determined  by the regularisation parameter $\varepsilon$. Note that this $\varepsilon$ is not a physical parameter as it has no real life significance, i.e. it is merely introduced to make Coulomb's friction law non-singular at the bonded region.\\

In Fig. \ref{Oden1Eg2}, if the applied traction at $x^2=\frac{1}{2}\pi$ is larger than the traction at $x^2=-\frac{1}{2}\pi$ (i.e. if $\tau_\text{max} > \tau_0$), then it will no longer have a symmetric profile in the azimuthal direction, and for this case the right side (i.e. where slip $= 1$) will be much greater than the left side (i.e. where slip $= -1$). Thus, if one increases the traction at $x^2=\frac{1}{2}\pi$ enough, then the entire body will slip in the positive azimuthal direction, i.e. slip $= 1$,  $\forall~\theta$. Therefore, given a fixed coefficient of friction (i.e. for fixed $\nu_F$), we can now determine that what is the minimum tension ratio (i.e. $\delta \tau = \tau_\text{max}/ \tau_0$) that is required for the elastic body to attain limiting-equilibrium. Which in turn help us to understand that for a given a tension ratio, what is the minimum coefficient of friction required for the elastic body to attain limiting-equilibrium. Thus, the subject of the remainder of this section is to investigate how this slip region behaves for a given set of variables, as this will infer the relationship between the two models for Coulomb's law of static friction. \\

\begin{figure}[!h]
\centering
\begin{subfigure}{.5\linewidth}
    \centering
    \includegraphics[width=1\linewidth]{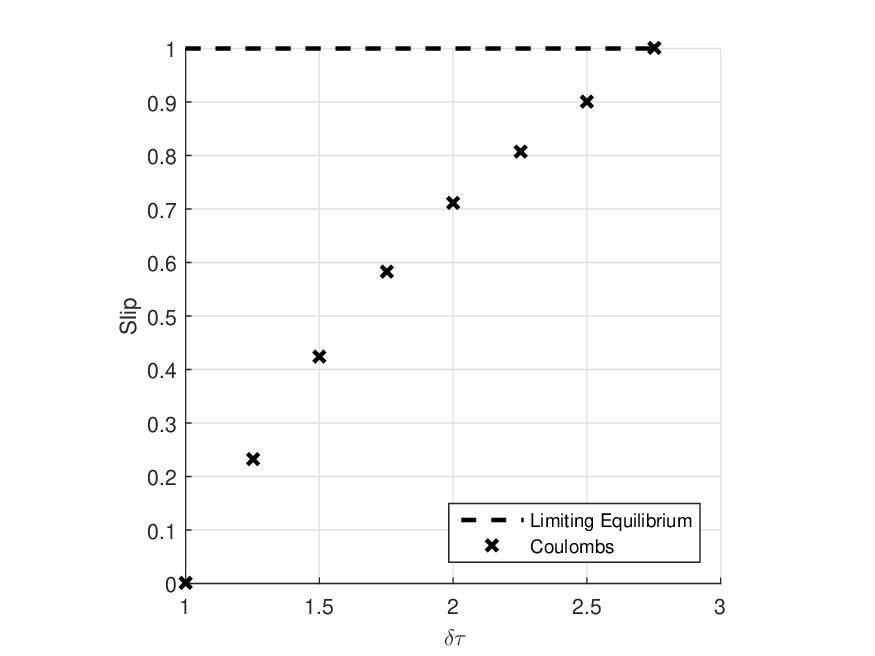}
\caption{Slip for $\delta\tau$\label{Oden1T}}
\end{subfigure}%
\begin{subfigure}{.5\linewidth}
    \centering
    \includegraphics[width=1\linewidth]{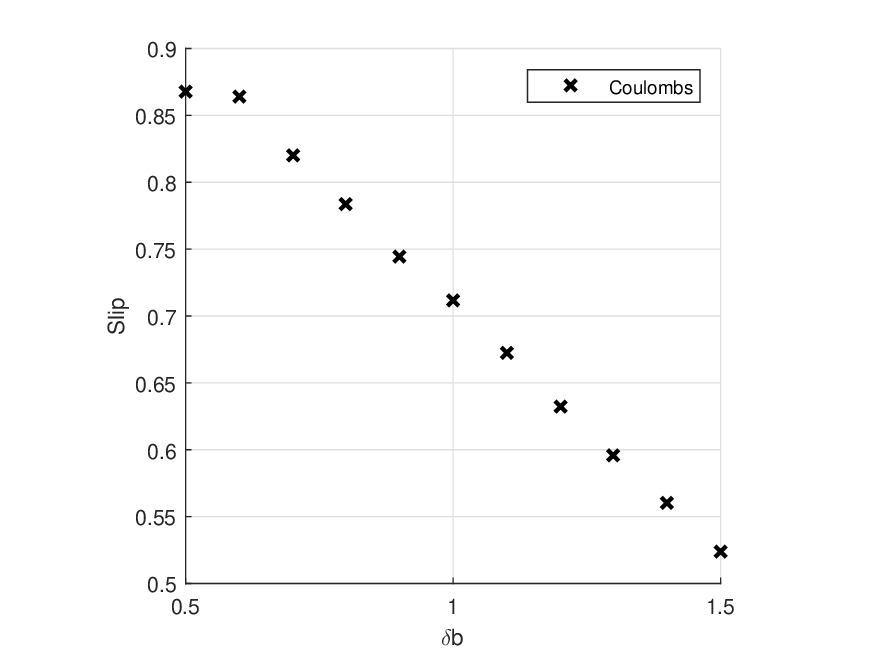}
\caption{Slip  for $\delta b$\label{OdenCur}}
\end{subfigure}
\begin{subfigure}{.5\textwidth}
    \centering
    \includegraphics[width=1\linewidth]{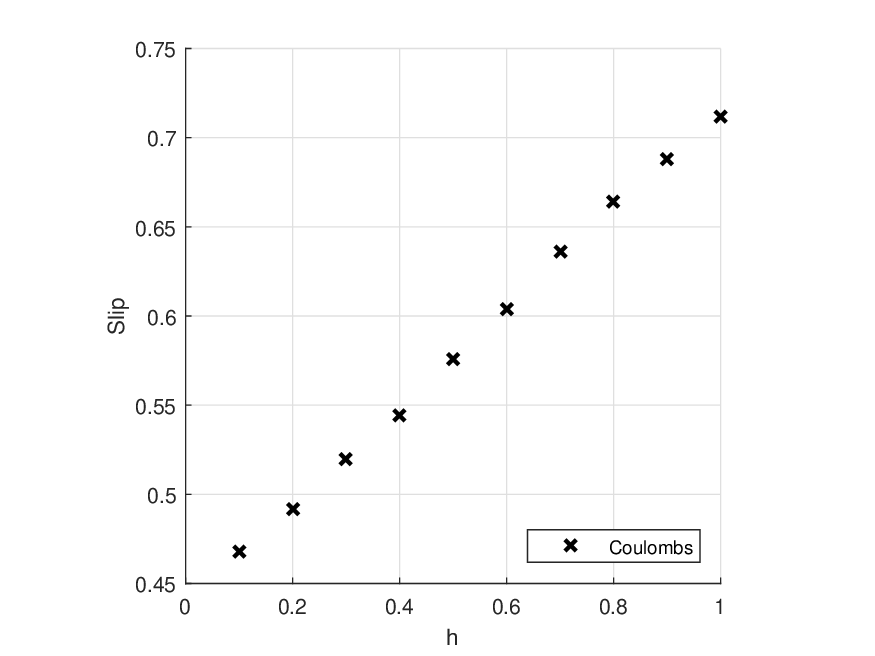}
\caption{Slip for $h$\label{OdenH}}
\end{subfigure}%
\begin{subfigure}{.5\textwidth}
    \centering
    \includegraphics[width=1\linewidth]{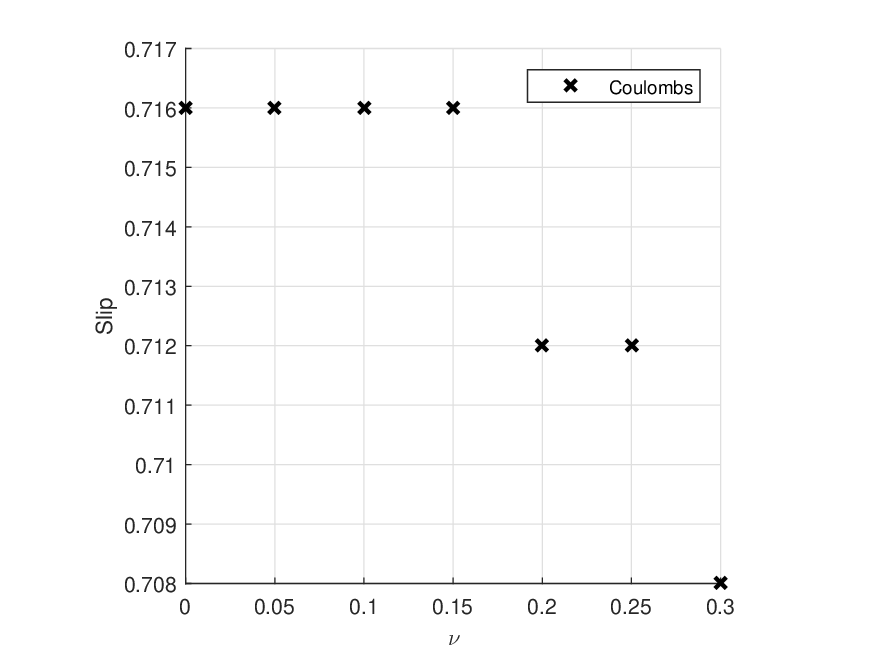}
\caption{Slip for $\nu$\label{OdenPoi}}
\end{subfigure}
\caption{Kikuchi and Oden's model Coulomb's law of static friction's (key: Coulombs) Slip profiles for varying $\delta \tau$, $\delta b$, $h$ and $\nu$}
\label{fullset}
\end{figure}

Fig. \ref{fullset} shows the Slip profile for varying $\delta \tau$, $\delta b$, $h$ and $\nu$, where the Slip is calculated by taking the mean value of all the slip and stick points at the contact boundary (i.e. at $\overline{\omega}^{\text{New}}$). Recall that at each grid point at the contact boundary, positive $x^2$ directional slip, negative $x^2$ directional slip and bonding (i.e. stick) has the values $1$, $-1$ and $0$ respectively. Thus, we define Slip more precisely as follows
\begin{align*}
\text{Slip} = \frac{\text{sum of all stick and slip values at grid points at the contact boundary}}{\text{number of grid points at the contact boundary}}~.
\end{align*}
Slip = $1$ implies that the tension ratio (i.e. $\delta \tau$) is large enough for the elastic body to be at limiting-equilibrium, i.e. the elastic body is at the point of debonding. Recall that the capstan equation is valid when the elastic body at limiting-equilibrium (here, we are neglecting the steady-equilibrium case). Thus, by feeding our tension ratio for the Slip = $1$ case in to the capstan equation, we can obtain a value for the coefficient of friction implied by the capstan model. Fig. \ref{Oden1T} shows how close the elastic body is to limiting-equilibrium for varying tension ratios. We see that as the tension ratio increases, the elastic body gets closer to fully debonding. In fact, when $\delta \tau>2.75$, our elastic body is fully debonded from the rigid obstacle and sliding in the positive $x^2$ direction. To compare it against the capstan equation, we invoke corollary \ref{crlPrism} with $\delta\tau=2.75$, which indicates that the coefficient of friction implied by the capstan equation is $\mu_F = 0.322$, regardless of the Poisson's ratio of the elastic body. This result shows that the coefficient of friction implied by the capstan equation is an underestimate of  the coefficient of friction implied by Kikuchi and Oden's model, i.e. $\mu_F\leq\nu_F$. Note that Martin  and Mittelmann \cite{martin194618} also demonstrate that the coefficient of friction implied by capstan equation is an underestimate of the coefficient of friction implied by their experiments on wool fiber measurements.\\

Fig. \ref{OdenCur} shows how close the elastic body is to limiting-equilibrium for varying radii ratio, $\delta b$, of the contact region. We see that as the $\delta b$ increases, the modified Kikuchi and Oden's model moves away from  limiting-equilibrium. For the contact interval $(-\frac12\pi+\varepsilon,\frac12\pi-\varepsilon)$ where $\varepsilon>0$, this observation can be interpreted as, as the curvature of the contact region increases, modified Kikuchi and Oden's model moves away from  limiting-equilibrium and this observation coincides with capstan equation for an elliptical prism case (\ref{PrismEqn}). Note however, for the contact interval $[-\frac12\pi,\frac12\pi]$, tension ratio of the capstan equation for an elliptical prism case (\ref{PrismEqn}) is invariant with respect to $\delta b$.\\

Fig. \ref{OdenH} shows how close the elastic body is to limiting-equilibrium for varying thicknesses of the overlying body. We see that as $h$ decreases, the modified Kikuchi and Oden's model moves away from limiting-equilibrium, i.e. as the thickness decreases, Kikuchi and Oden's model behaves more like the capstan equation. This possibly implies a convergence between both models in the limit $h \to 0$. However, the force predicted by modified Kikuchi and Oden's model is still an underestimate to what is predicted by the Kikuchi and Oden's model capstan equation.\\

Another numerical experiment that we conduct examines the behaviour of the body under varying Young's modulus, $E$, and for this experiment we consider  $E \in[500\text{Pa},1500\text{Pa}]$. We find that the body has a constant slip of $71.2\%$ in the positive $x^2$ direction for all values of Young's moduli. This is intuitive as, whatever the value of Young's modulus is (given that it is not zero or infinite), one can always rescale the displacement field without affecting the final form of the solution. This result also tend to coincide with the capstan equation, as the capstan equation is also invariant with respect to the Young's modulus of the elastic body.\\

Fig. \ref{OdenPoi} shows how close to the elastic body is to  limiting-equilibrium for varying Poisson's ratio. We see that as $\nu$ increases, the modified Kikuchi and Oden's model moves away from the limiting equilibrium, i.e. as the body becomes incompressible, one needs to apply more force to debond the body from the rigid surface. This effect is small in comparison to other results, but it is still numerically observable. This is a surprising result as this tends to contradict  the special case of capstan model that is compatible with non-zero Poisson's ratios  (i.e. corollary \ref{crlPrism}) as this model is invariant with respect to Poisson's ratio of the elastic body.

\section{Conclusions}

In our analysis, we have taken Coulomb's formulation of Amontons' laws of dry friction  (\ref{FrictionLaw01}) and extend it to model thin elastic bodies on rough rigid surfaces in an attempt to extend the capstan equation (\ref{CapstanEqn}) to more general geometries. First, we derived closed form solutions for a membrane with a  zero Poisson's ratio (or a string with an arbitrary Poisson's ratio) supported by a rigid prism at limiting-equilibrium (static-equilibrium case and steady-equilibrium case), and supported by a rigid cone at limiting-equilibrium (static-equilibrium case only). The solutions that are derived from our models indicate that the stress profile of the elastic body depends on the change in curvature of the rigid obstacle at the contact region.\\

As a comparison, we adapted Kikuchi and Oden's  model for Coulomb's law of static friction \cite{Kikuchi} to model curvilinear elastic bodies. For this model, we conducted numerical experiments to observe the relationship between between Kikuchi and Oden's  model Coulomb's law of static friction in curvilinear coordinates and our generalised capstan equation. To do this, we modelled an elastic body over a rigid rough prism whose cross section is elliptical. Our numerical results indicate the following: for a fixed coefficient of friction (i) for a given tension ratio, the coefficient of friction implied by the capstan equation is an underestimate of the coefficient of friction implied by Kikuchi and Oden's  model, i.e. $\mu_F\leq\nu_F$; (ii) as the curvature of the contact region increases, one requires a larger force to debond the body, which is a result that coincides with the modified capstan equation; (iii) as the thickness of the body decreases, one require a larger force to debond the body; however, this force is still an underestimate to what is predicted by the modified capstan equation; (iv) Young's modulus of the body does not affect the governing equations at the contact region, which is another result that coincides with the capstan equation (modified or otherwise); and (v) as Poisson's ratio of the body increases, one requires a larger force to debond the body. The last result implies that an incompressible elastic body, such as rubber, tend to be more difficult to debond from a rigid surface relative to a compressible body with the same coefficient of friction acting on the contact region. Now, this latter result is surprising as this behaviour cannot  be predicted by the special case of the modified capstan equation (a capstan model that is compatible with non-zero Poisson's ratios) as this model is invariant with respect to Poisson's ratio of the elastic body.\\

The conclusion inferred by our numerical modelling can be summarised as follows: for the same value of coefficient of friction, different models predict vastly different friction profiles (e.g. different limiting-equilibriums), which, in turn, implies that calculating the coefficient of friction from a tension pair $\{T_0, T_\text{max}\}$ is model dependent, i.e. the implied-coefficient of friction is model dependent. Thus, our analysis shows that modelling friction is not a well understood problem as different models (that supposedly model the same physical phenomena with common roots stretching back to the Amontons' laws of friction) predict different outcomes.\\

\section*{Acknowledgments}
We thank Dr Nick Ovenden (UCL) and Prof Alan Cottenden (UCL) for their supervision, Brad Turner (TEKOR) for his assistance, and Christopher Law for the illustrations.

\bibliographystyle{./model1-num-names}
\bibliography{GeneralisedCapstanEquation}%
\biboptions{sort&compress}

\end{document}